\documentclass[pra,aps,superscriptaddress,twocolumn,floatfix,notitlepage,longbibliography]{revtex4-2}

\usepackage{amsmath}
\usepackage{amsfonts}
\usepackage{amssymb}
\usepackage{wasysym}
\usepackage{bm}
\usepackage{mathrsfs}
\usepackage{graphicx}
\usepackage{color}
\usepackage{xcolor}
\usepackage{xspace}
\usepackage{stmaryrd}
\usepackage{laussy}
\usepackage{url}
\usepackage[sort&compress]{natbib}
\usepackage{multibib}
\newcites{supp}{Supplementary References}

\usepackage[colorlinks, linkcolor=magenta, citecolor=magenta,
urlcolor=blue]{hyperref}
\begin{document} 
\title{Bosons from classical states are not quantum correlated}
\title{Boson correlations are spurious for classical states}

\author{Daniel E.~Salazar}
\author{Fabrice P.~Laussy}
\email{fabrice.laussy@gmail.com}
\affiliation{Instituto de Ciencia de Materiales de Madrid ICMM-CSIC, 28049 Madrid, Spain}

\date{\today}

\begin{abstract}
  We show that boson correlations from quantum states with a
  Glauber-Sudarshan representation of their density matrix which
  provides a well-behaved probability distribution---including
  coherent states, thermal states, and all states that can be deemed
  classical---are a manifestation of the Simpson paradox: they are
  spurious correlations from statistical (ensemble) averages over
  uncorrelated measurements made in varying geometries, due to a
  process of symmetry-breaking as a confounding factor.  Bosonic
  correlations encoded by the wavefunction appear to be formed in the
  geometry assumed, which however is not that of the statistical
  ensemble but varies from realization to realization. This calls to
  distinguish between quantum and statistical averages and sheds new
  understandings on the fundamental problems of nonclassicality and
  quantum advantage. 
\end{abstract}

\maketitle

Quantum mechanics is a statistical theory and statistics is well-known
to be prone to paradoxes and to produce mystifying, counter-intuitive
correlations~\cite{pawitan_book24a}. One of the earliest commotion in
quantum weirdness is Hanbury Brown and Twiss's report that independent
photons are correlated upon detection~\cite{hanburybrown56a}. In
Hanbury Brown's own words: ``to a surprising number of people the idea
that the arrival of photons at two separated detectors can ever be
correlated was not only heretical but patently absurd, and they told
us so in no uncertain terms, in person, by letter, in print, and by
publishing the results of laboratory experiments, which claimed to
show that we were wrong''. Purcell advocated that ``the Brown--Twiss
effect, far from requiring a revision of quantum mechanics, is an
instructive illustration of its elementary
principles.''~\cite{purcell56a} Photons are bosons and boson
correlations emerge from the symmetrization of their wavefunction: the
lack of knowledge of which detected boson corresponds to which
underlying production process requires to consider both alternative on
equal footing, indeed, in a superposition. Such superpositions can be
interfered, leading to ideas of quantum
computation~\cite{deutsch85a}. Glauber formalized the ideas of
multi-order correlations from quantum states of the light field into a
theory of quantum optical coherence~\cite{glauber63b} that introduces
the coherent
state~$\ket{\alpha}\equiv\exp(-|\alpha|^2/2)\sum_{n=0}^\infty{\alpha^n\over\sqrt{n!}}\ket{n}$
as the one with no correlations in the sense of Hanbury Brown and
Twiss: photons from a coherent states are detected
independently. Consequently, all their $n$-order correlation functions
factorize as products of the first-order one. A coherent state can
therefore be seen as the quantum description of a classical state.
Other states, such as the thermal state which describes photons from
stars or light bulbs---with which Hanbury Brown demonstrated his
effect~\cite{hanburybrown56b}---can be expressed in the basis of
coherent states, through the so-called Glauber--Sudarshan~$P$
representation of the density matrix
$\rho=\int P(\alpha,\alpha^*)\ketbra{\alpha}{\alpha}\,d^2\alpha$.
This led to a controversy between Glauber~\cite{glauber63c} and
Sudarshan~\cite{sudarshan63a} backed-up by Mandel and
Wolf~\cite{mandel63a}. The latter authors regarded the~$P$ function as
a tool to decorrelate photons (a view which we shall materialize in
the following) and make a ``complete equivalence'' between the
classical and quantum mechanical approaches, while Glauber took the
view that the~$P$ function is ``an intrinsically quantum-mechanical
structure and not derivable from classical
arguments''~\cite{glauber63c}. We show here how bringing such
discussions to multiphoton and multimode correlations clarifies this
dispute. Specifically, we show that, for states that can be regarded
as ``classical''---in the sense that their~$P$ distribution is a
well-behaved distribution function---any correlation in multiphoton
observables can be understood as the result of an aggregation paradox.
We find that the sampling of particles is made independently, as for
classical particles, and from one and the same probability
distribution, but which itself varies from measurement to measurement,
according to the quantum state~$P$ and a fundamental process of
symmetry breaking. This is only because observables are construed from
an ensemble of different distributions, that correlations
\emph{appear} to be produced. In this sense, bosonic correlations from
classical states are an artifact from biased sampling. This is in
stark contrast with the traditional association to bosonic
correlations inherent to the symmetry of the wavefunction.  Quantum
states which have no classical probability distribution---such as Fock
states, or even Gaussian states such as squeezed states---still give
rise to genuine bosonic correlations that can power quantum
information processing.

\begin{figure*}
  \centering
  \includegraphics[width=\linewidth]{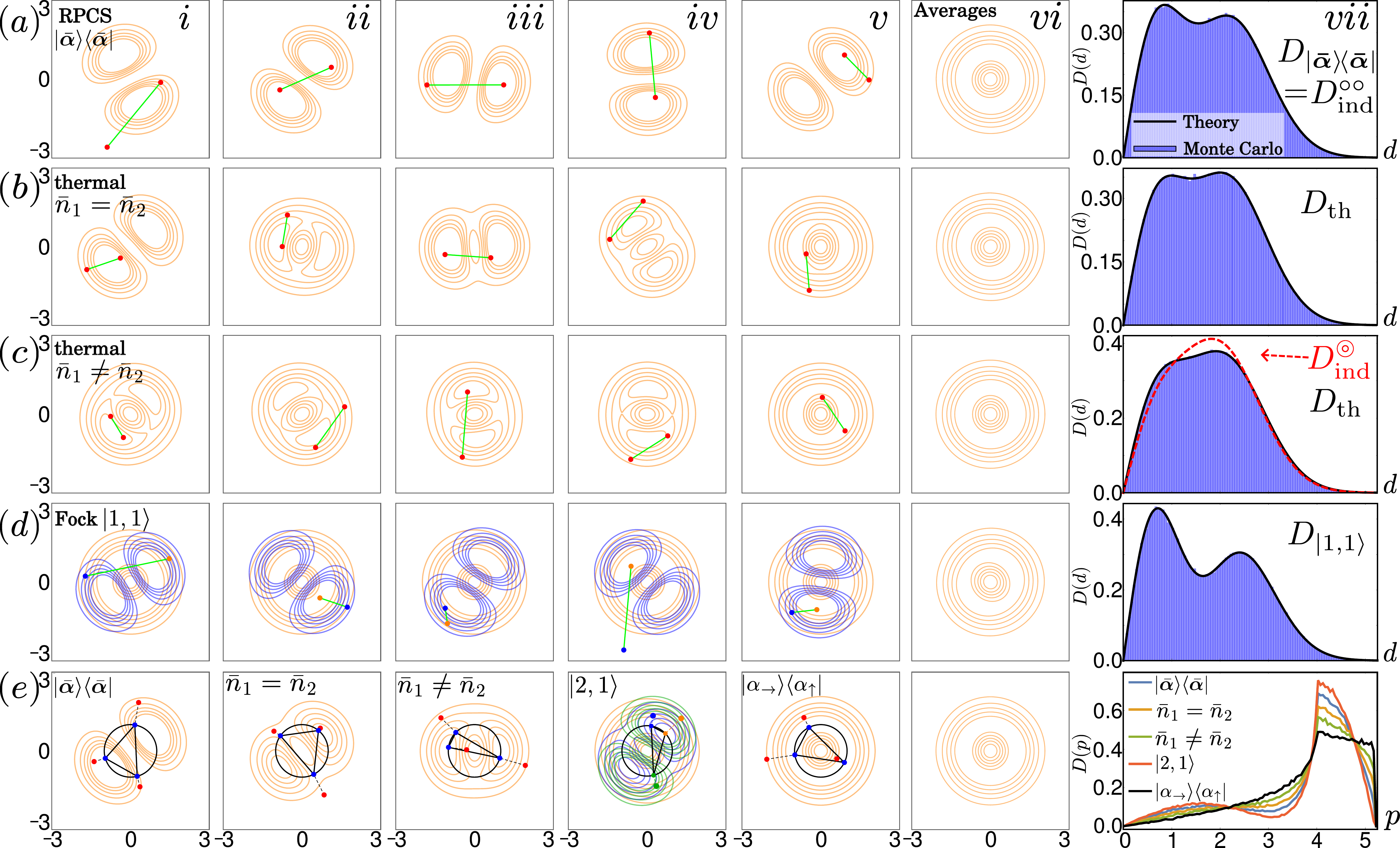}
  \caption{Illustration of our main statement through Monte-Carlo
    samplings for various families of quantum states: (a) RPCS, (b)
    Thermal states with (b) same population for both modes and with
    (c) $3.5$~times more particles in one branch than the other, and
    (d) Fock states with exactly one particle in each mode. The
    columns $i$--$v$ show random sampling for those cases from the
    underlying distributions shown as countour plots: RPCS are perfect
    dipoles but with random orientation, thermal states are distorted
    dipoles and Fock states require correlated geometries for the
    sampling, implying genuine quantum correlations. In contrast, in
    cases (a--c), particles are sampled independently from the same
    configuration, showing that they are intrinsically \emph{not}
    correlated.  Since the final geometry reproduced by all particles
    from all frames, in columns $vi$, is the same regardless of the
    underlying quantum state, the illusion of correlation is
    tenacious. Column~$vii$ shows a possible observable for such
    measurements: absolute distances~$d$ for two bosons and
    perimeters~$p$ of triangles projected on the circle for three
    photons.}
  \label{fig:FriMar27013221PMCET2026}
\end{figure*}

We articulate our discussion with correlations in space, rather than
in time, since the additional degrees of freedom bring more room to
study the phenomenon, and allow for a transparent reading of the
nature and meaning of the correlations. Specifically, we rely on the
formalism by Zubizarreta \emph{et
  al.}~\cite{arXiv_zubizarretacasalengua24a} to interfere two vortices
formed by LG$_{p=0}^{\ell=\pm1}$ beams superimposed in different
quantum states. This can be generalized to the~$N$-body wavefunction
$\rho^{(N)}(\mathbf{r}_1,\dots,\mathbf{r}_N) =
\langle:\hat{n}^{(1)}(\mathbf{r}_1)\cdots\hat{n}^{(1)}(\mathbf{r}_N):\rangle~$
built from the one-body density operators
$\hat{n}^{(1)}(\mathbf{r}) =
\hat{\Psi}^\dagger(\mathbf{r})\hat{\Psi}(\mathbf{r})$ where the
quantum field operator $\hat{\Psi}(x)\equiv\sum_m \phi_m \hat{a}_m$ is
expanded on a set of basis eigenfunction~$\phi_m$ attached to their
corresponding annihilation operator~$\hat a_m$ with bosonic
algebra~$[a_m,\ud{a_l}]=\delta_{m,l}$. For simplicity and with little
loss of generality, we consider two modes $a_1$ and $a_2$ only, which
we shall take to be two vortices, so that we can focus on the angular
part of the wavefunction~$\Theta(\theta_1,\cdots,\theta_N)$ since the
radial part $R(r_1,\cdots,r_N)\equiv\prod_{i=1}^N {2r_i^2e^{-r_i^2}}$
features no correlations. The angular part, on the other hand, is a
combinatorial superposition over the set $\mathfrak{S}[S_k^N]$ of the
${N\choose k}$ permutations of the~$\ell=\pm 1$ sign of the phase
circulation, which leads to the normalized (denoted
with~{\raisebox{-1.5ex}{\Large\textasciitilde}}) $N$-body angular
wavefunction:
\begin{multline}
  \label{eq:SatApr4112658AMCEST2026}
  {\widetilde\Theta}^{(N)}(\theta_1\dots, \theta_N) = \frac{1}{(2\pi)^N{\sum^n_{k=0}C_{k,n-k}\binom{n}{k}}}\times\\
  {\sum^{n}_{k=0}C_{k,n-k}}\sum_{\sigma_1,\sigma_2\in\mathfrak{S}[S^n_k]}\cos\big(\sum^n_i\sigma_1(i)\theta_i-\sum^n_j\sigma_2(j)\theta_j\big)
\end{multline}
where~$C_{ij}\equiv\langle a^{\dagger i}_1 a_1^ia_{2}^{\dagger
  j}a_2^j\rangle$ captures the joint quantum state of the two
vortices. For two particles only,
Eq.~(\ref{eq:SatApr4112658AMCEST2026}) reduces to the
rotation-invariant ($\Delta\theta\equiv\theta_2-\theta_1$) expression
$\widetilde{\Theta}^{(2)}(\theta_1,\theta_2) =
\frac{1}{(2\pi)^2}\left(1+2\mathscr{C}\cos(2\Delta\theta)\right)$,
where $\mathscr{C}\equiv{C_{11}}\big/({C_{20}+2C_{11}+C_{02}})$ for
general quantum states, evaluates to
$\mathscr{C}_{\ket{n_1n_2}}\equiv{n_1n_2}/[(n_1+n_2)(n_1+n_2-1)]$ for
Fock states~$\ket{n_1n_2}$ with exactly~$n_i$ particles in
mode~$i=1,2$,
$\mathscr{C}_{{\bar n_1,\bar n_2}}\equiv{1\over2}{\bar n_1\bar
  n_2}/{(\bar n_1^2+\bar n_1\bar n_2+\bar n_2^2)}$ for thermal states
$\rho_\mathrm{th}={1\over(1+\bar n_1)(1+\bar
  n_2)}\sum_{\mu,\nu=0}^\infty\big({\bar n_1\over\bar
  n_1+1}\big)^\mu\big({\bar n_2\over\bar
  n_2+1}\big)^\nu\ketbra{\mu\nu}{\mu\nu}$ and
$\mathscr{C}_{{\ketbra{\boldsymbol{\bar\alpha}}{\boldsymbol{\bar\alpha}}}}\equiv{|\alpha_1|^2|\alpha_2|^2}/{(|\alpha_1|^2+|\alpha_2|^2)^2}$
for Random-Phase Coherent States (RPCS)
where~$\boldsymbol\alpha\equiv(\alpha_1,\alpha_2)^T$ and
$\rho_{\ketbra{\boldsymbol{\bar\alpha}}{\boldsymbol{\bar\alpha}}}=e^{-|\alpha_1|^2-|\alpha_2|^2}\sum_{\mu,\nu=0}^\infty{\alpha_1^\mu\alpha_2^\nu\over\mu!\nu!}\ketbra{\mu\nu}{\mu\nu}$. Such
a symmetrization of the wavefunction results in correlations from
various quantum states for the vortices, whose magnitudes are
modulated by~$\mathscr{C}$, with Fock states being the most
correlated. Considering, for instance, distances between photons all
sampled from the same donut shape (column~$vi$ of
Fig.~\ref{fig:FriMar27013221PMCET2026}) Zubizarreta \emph{et
  al}~\cite{arXiv_zubizarretacasalengua24a} report bimodal
distributions of distances, shown in column $vii$ (rightmost). This
structure is most pronounced for the Fock state in panel d--$vii$ but
remains visible for all quantum states, at the exception of
uncorrelated (classical) particles, sampled independently from the
donut. In this case, shown as the dashed red line in Panel~(c)--$vii$,
$D_\mathrm{ind}^\circledcirc(d)\equiv{d\over16}(8+d^4)e^{-d^2\!/2}$
has only one maximum (``ind'' is for independent). The departures, due
to correlations, are attributed to boson coalescence, whereby the two
particles tend to be detected at the same point or, equally likely by
symmetry, on opposite sides of the donut. Thermal light is merely less
correlated than Fock states, but the bunching and cluttering
tendencies remain.  Importantly, coherent states are \emph{not}
correlated in this way, and although their wavefunction is also
symmetric under exchanges of particles, the coherent superpositions
with Poisson weight result in a cancellation of such correlations, in
the sense that
$D_\mathrm{\ket{\alpha_\rightarrow}\ket{\alpha_\uparrow}}(d)=D_\mathrm{ind}^\circledcirc(d)$
is the distribution of independent particles. Note that due to phase
locking, one must superimpose dipoles instead of vortices to form a
vortex and sample from the donut shape. Particles sampled from this
coherent-state donut are not correlated. Similarly, if one would
sample from $\ket{\alpha_\la}\ket{\alpha_\ra}$, one would obtain a
dipole whose orientation is fixed by the relative phase
$\operatorname{arg}(\alpha_\la/\alpha_\ra)$, but whose particles would
produce the same observables as if they were sampled independently,
which, for sampling from a dipole (not a donut)
gives~$D_{\mathrm{ind}}^{\circ\circ}\equiv{d\over32}(24-8d^2+3d^4)e^{-d^2\!/2}$. All
other quantum states produce correlations on their assumed geometry,
which come from their respective two-body density matrices
$\widetilde\rho^{(2)}(\mathbf r_1,\mathbf r_2) \equiv
\widetilde\Theta^{(2)}(\theta_1,\theta_2) \prod_{j=1}^2 |R_1(r_j)|^2$,
i.e., for distances, from
$D(d) = \int\widetilde\rho^{(2)}(\mathbf r_1,\mathbf
r_2)\,\delta\!\left(d-|\mathbf r_1-\mathbf r_2|\right)d^2\mathbf r_1\,
d^2\mathbf r_2$.  For instance,
$D_{\ket{1,1}}(d)={d\over8}(8-4d^2+d^4)e^{-d^2\!/2}$.  Notably,
sampling RPCS from a donut provides stronger correlations than
sampling thermal states from the same donut, with
distribution~$D_\mathrm{th}(d)={d\over12}(8-2d^2+d^4)e^{-d^2\!/2}$ in
panel~c--$vii$, as compared to the distribution
$D_{{\ketbra{\boldsymbol{\bar\alpha}}{\boldsymbol{\bar\alpha}}}}={d\over32}(24-8d^2+3d^4)e^{-d^2\!/2}$
in~a--$vii$.  Despite its association to classical light, the RPCS is
a non-Gaussian state, since the only diagonal state that is Gaussian
is the thermal state~\cite{xu16b}. Indeed, one could expect
non-Gaussian states to feature stronger correlations than Gaussian
ones. However, we are about to reconsider such qualifications.
Non-Gaussianity does not refer to non-classicality since the~$P$
distribution for RPCS has a classical interpretation, namely,
$P_{{\ketbra{\bar\alpha_0}{\bar\alpha_0}}}(\alpha)=\delta(|\alpha|-|\alpha_0|)$
i.e., a state of fixed intensity but random phase (as the name
posits). The same happens for thermal states whose $P$ function is a
Gaussian
$P_{\bar n}(\alpha)=\exp(-|\alpha^2|/\bar n)/\pi$. We
now show that such states---with a classically-behaved $P$
function---are not correlated in the same way as Fock states but as a
result of statistical averages over uncorrelated sampling. Namely, we
prove in the Supplementary Material~\cite{bosoc_sup} that the
many-body density matrix, for any~$N\in\mathbb{N}$, of two vortices
whose quantum states are RPCS---thus with joint quantum
state~$P_{\ketbra{\boldsymbol{\bar\alpha}}{\boldsymbol{\bar\alpha}}}(\alpha,\alpha')=P^{\la}_{{\ketbra{\bar\alpha_1}{\bar\alpha_1}}}(\alpha)P^{\ra}_{{\ketbra{\bar\alpha_2}{\bar\alpha_2}}}(\alpha')$---is obtained as
a phase-average product of $N$ one-body wavefunctions:
\begin{equation}
  \label{eq:TueMar31043938PMCEST2026}
  \widetilde\Theta^{(N)}_{\ketbra{\boldsymbol{\bar\alpha}}{\boldsymbol{\bar\alpha}}}(\theta_1,\cdots,\theta_N)={1\over2\pi}\int_0^{2\pi}
  \prod_{j=1}^N\widetilde\Theta_{\boldsymbol{\alpha}}^{(1)}(\theta_j+\eta)\,d\eta\,,
\end{equation}
where
$2\pi\widetilde\Theta_{\boldsymbol{\alpha}}^{(1)}(\theta)\equiv\big||\alpha_1|e^{i\theta}+|\alpha_2|e^{-i\theta}\big|^2/(|\alpha_1|^2+|\alpha_2|^2)=1+2\frac{|\alpha_1\alpha_2|}{|\alpha_1|^2+|\alpha_2|^2}\cos(2\theta)$
is the one-body probability distribution for the angles of a distorted
dipole (exact dipole when~$|\alpha_1|=|\alpha_2|)$ of
orientation~$\eta\equiv\operatorname{arg}(\alpha_1/\alpha_2)$.  For
each fixed reference angle~$\eta$---that defines the origin of angles
for all other~$\theta_j$---each particle in the integrand on the rhs
is sampled from the same
distribution~$\widetilde\Theta^{(1)}_{\boldsymbol{\alpha}}$ and,
crucially, independently from all the other particles, which is what
the product expresses. Fixing~$\eta$, one obtains the uncorrelated
case of classical particles sampled from a dipole of said orientation
(radii remain sampled independently). At this stage, the theory stops
discriminating between two types of averages: one is the quantum
average whereby the integral in
Eq.~(\ref{eq:TueMar31043938PMCEST2026}) is a quantum superposition,
whereby all the possible phases are carried along in the state, while
the other is the statistical average to be performed over an ensemble
of independent cases, each of which is a random~$\eta$ realization
that occurs through the well-known mechanism of symmetry breaking. In
the latter case, the system chooses randomly a phase, that sets the
origin~$\eta$ and therefore a dipole of corresponding
orientation. From this dipole, $N$ particles are sampled
independently, from which any~$N$-photon observables can be
obtained. Repeating the experiment, one builds the statistical
observable.

Our main observation then reads as follows: in the scenario which we
have just explained, and that is illustrated with numerical Monte
Carlo simulations in Fig.~\ref{fig:FriMar27013221PMCET2026},
\emph{there are no boson or quantum correlations} of any type between
particles from states which have a positive-definite $P$
representation, such as thermal states, despite their observed
bunching. This arises due to the assumption that those are sampled
from a given geometry---in our case, a donut as shown in the
column~$vi$ of averages---whereas they are truly sampled from another
geometry---a more or less distorted dipole depending on the quantum
state, of varying orientation---which remains hidden at the time of
their uncorrelated sampling. Particular cases sampled from their
respective~$P$ functions are shown in columns~$i$--$v$. 
The same applies to any quantum state which has a
classical-probability interpretation for the~$P$ function: bosons
sampled from such states are not correlated, but appear so as a result
of assuming that they are sampled from the same distribution. The
illusion is a good one, because all the various quantum states define
the same ``final geometry'' from which one assumes the sampling
(again, the donut), but at the time of the actual measurement, the
genuine geometry differs from one measurement to the next, which is
why a thermal state, a RPCS or a coherent state exhibit different
``apparent'' correlations: they are uncorrelated on genuine but
different geometries, and thus appear correlated on the same apparent,
but misleading, geometry.
The strong correlations from RPCS can now be understood as originating
from each measurement being made on a dipole, and, therefore,
$D_{{\ketbra{\boldsymbol{\bar\alpha}}{\boldsymbol{\bar\alpha}}}}=D_{\mathrm{ind}}^{\circ\circ}$. These
are the strongest correlations that one can obtain from independent
(classical) sampling.  For the thermal state---the most widespread but
also the most important case, including for its historical role in the
development of quantum optics---its $P$ function is a Gaussian state
$P_{\mathrm{th}}(\alpha_1,\alpha_2) = \frac{1}{\pi \bar n_1\bar
  n_2}e^{-(|\alpha_1|^2/\bar n_1)-(|\alpha_2|^2/\bar n_2)}$. In the
case~$\bar n_1=\bar n_2$, the one-particle wavefunction from which
independent particles are to be sampled, takes the form
$\widetilde\Theta_t^{(1)}(\theta)\equiv 1+2\sqrt{t(1-t)}\cos(2\theta)$
where~$0\le t\le 1$~\cite{bosoc_sup}. The geometry from which the
particles are sampled in each measurement, is now distorted
differently on each realization, in addition of its random
orientation. 
Consequently, the same procedure as above leads us to the
de-correlation of bosons now expressed as
$\widetilde\Theta_\mathrm{th}^{(N)}(\theta_1,\cdots,\theta_N)=
\frac{1}{2\pi}\int_0^1 \int_0^{2\pi} \prod_{j=1}^N
\widetilde{\Theta}^{(1)}_{t}(\theta_j+\eta)\,d\eta\,dt$,
which is the counterpart of Eq.~(\ref{eq:TueMar31043938PMCEST2026})
but now each sampling is done on a greatly varying geometry, as shown
in Fig.~\ref{fig:FriMar27013221PMCET2026}(b). 
Such distortions cancel out into the perfect donut shape when averaged
over, from which one has the illusion of sampling correlated
particles, whereas in effect, independent particles were sampled from
varying cases.  There is a remarkable and curious property that the
correlations which we have discussed so far remain independent of the
intensity of the fields, meaning that one can increase the signal
merely by scaling up the populations of the modes, allowing us to
easily measure high-order correlations. This remains true for RPCS
even if the two modes~$\la$ and~$\ra$ have different intensities,
$|\alpha_1|^2\neq|\alpha_2|^2$, but for thermal state, this is true
only if~$\bar n_1=\bar n_2$, which is the case we have considered up
to now. If not, aspects of normalization which we discuss in the
Supplementary Material~\cite{bosoc_sup} lead to weighting the
observables in a way that needs to be taken into account, but that
cancel for equal populations. This correction is simple enough,
though, it consists in sampling from the normalized
distribution~$(|\alpha_1|^2+|\alpha_2|^2)^2P_\mathrm{th}(\alpha_1,\alpha_2)$. The
weight thus added now correlates the amplitudes of the two modes that
define the sampling geometry, but this merely constrains the
distribution of the latter, and once chosen the one-particle
geometries, all particles remain sampled independently from it. We
show in Fig.~\ref{fig:FriMar27013221PMCET2026}(c) such a thermal case
where~$\bar n_1\ll\bar n_2$. Since in such a case, each geometry is
closer to the average, the correlations are weakened. For unbalanced
RPCS, the same distorted dipole is sampled over, only varying in its
orientation, so that correlations are also weakened.

\begin{figure}[tb!]
  \centering
  \includegraphics[width=\linewidth]{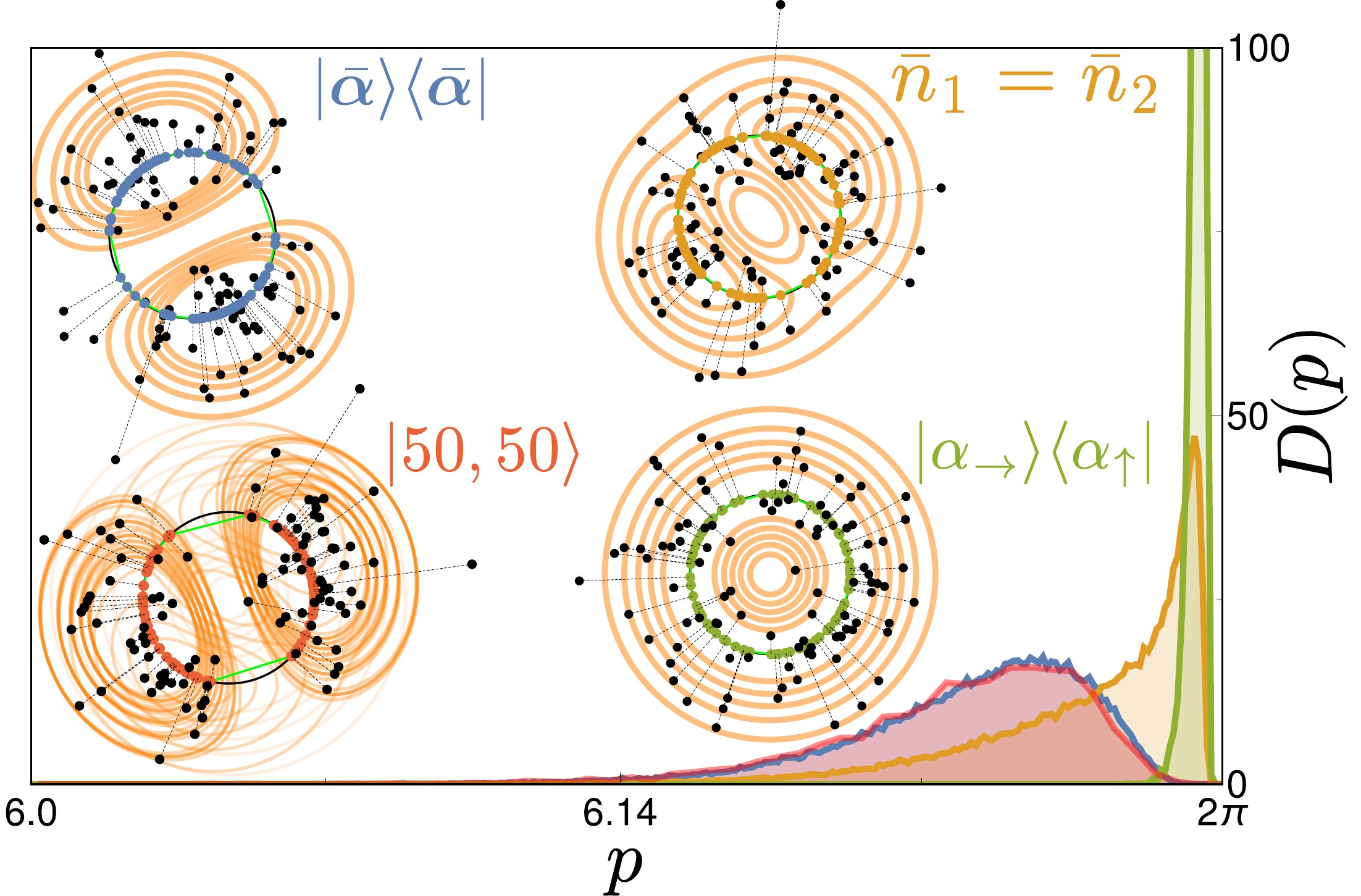}
  \caption{High multiphoton limit.  The distribution $D(p)$ of the
    perimeter of the polygon formed by projecting the points on the
    vortex's radius provides a 100-photon observable shown for a
    RPCS~$\ketbra{\bar\alpha}{\bar\alpha}$ (blue), thermal
    with~$\bar n_1=\bar n_2$ (orange), Fock~$\ket{50,50}$ (red) and
    classical
    (uncorrelated)~$\ket{\alpha_\rightarrow}\ket{\alpha_\uparrow}$
    cases.  Single realizations show how, with such a high number of
    particles, the underlying structure becomes directly visible in a
    single shot. The nature of boson correlations is revealed, as well
    as how multiphoton Fock states reduce to RPCS, to which their
    sampling distribution quickly converges to.}
  \label{fig:FriApr10094847AMCEST2026}
\end{figure}

The results which we have illustrated for spatial two-photon
correlations between two LG$^{\pm1}$ beams for a subset of popular
quantum states, are general and can be articulated for all quantum
states of more than two modes, in any basis (the dipolar basis is
discussed in the Supplementary Material~\cite{bosoc_sup}) as well as
for any number of particles. For illustrations, in
Fig.~\ref{fig:FriMar27013221PMCET2026}(e) we show the~$N=3$ case with
the distribution of perimeters of the points projected on the circle
(where correlations are embedded) for the quantum states previously
considered (more cases are shown in the Supplementary
Material~\cite{bosoc_sup}). Here too, positive-definite~$P$ lead to
sampling three independent particles from varying geometries. The same
applies to all experiments of this type. The case of high particle
numbers---illustrated for hundreds of particles in
Fig.~\ref{fig:FriApr10094847AMCEST2026}, still for perimeters on the
circle---also brings features of its own, that are beyond the scope of
our present discussion, but which we can summarize as follows: it
makes the symmetry breaking mechanism obvious in each single shot, as
there are now sufficiently enough points to image the geometry from
which particles are sampled (independently), and it calls for
contrasting various types of averages: statistical averages over
various repetitions, or quantum averages derived from sub-sampling
lower-order multiphoton observables from the pool made available by a
single realization of the experiment. We will discuss elsewhere how
those actually differ.


We have therefore shown that Boson correlations from states with a
positive-definite $P$ function are a manifestation of what is known in
statistics as the Simpson paradox~\cite{simpson51a}, where
correlations appear and typically reverse trend in aggregates as
compared to sub-groups. Our case is the particular one where no
correlations exist in subgroups but appear in their
aggregate~\cite{mittal91a}. This was first noted by
Pearson~\cite{pearson97a} who provides the example of uncorrelated
skull length and breadth within male and female subgroups, but when
the two groups are pooled, a substantial spurious correlation emerge.
Although Pearson developed statistics with biological problems in mind
(biometrics), he noted the relevance of this paradox for physics,
referring to spurious correlations between the judgments of different
observers, which reads like an eerily previous-century intuition of
qbism~\cite{fuchs14a}. While the ``paradox'' is easy to understand, in
Pearson's words: ``the fact that correlation can be produced between
two quite uncorrelated characters [\dots] must come rather as a
shock.'' This indeed has been the case in quantum optics.

Now that we have shown that a broad family of cases are, in fact,
completely classical and exhibit their phenomenology only as a result
of an artifact of an amalgamation paradox~\cite{good87a}, we must
hurry to restore genuine quantum correlations for states which have no
classical counterpart. This includes states for which the $P$ function
can be negative, such as thermal squeezed states or photon-added
coherent states, or even more pathological case where the function
becomes highly singular, such as Fock states or squeezed vacua, in
which cases it involves derivatives of the Dirac~$\delta$
function. For the most correlated quantum case---Fock states---the
sampling is not independent. The two-photon case is illustrated in
Fig.~\ref{fig:FriMar27013221PMCET2026}(d): a first photon is sampled
from the donut. This photon then pins the geometry for the second
photon, to a dipole whose orientation is given by the first
photon. The second photon is then sampled from this distribution,
which is both different from, and correlated with, the previous
one. The two photons are thus genuinely correlated. The case of three
photons is shown in Panel~(e--$iv$) of
Fig.~\ref{fig:FriMar27013221PMCET2026} and of 100~photons in
Fig.~\ref{fig:FriApr10094847AMCEST2026}. It is, however, striking that
already with three particles, the classical RPCS correlations are
already very close from those of the optimally quantum-correlated
case~$\ket{2_\la,1_\ra}$ (or $\ket{1_\la,2_\ra}$), and quickly become
indistinguishable as the number of particle increases (cf.~the blue
and red distributions in
Fig.~\ref{fig:FriApr10094847AMCEST2026}). This is because the sampling
geometry for each particle quickly converges to a stable one, thus
recovering the RPCS. This suggests new pictures for the quantum to
classical transition.

Our present discussion is important for many other reasons: first, it
brings a new angle to the thorny question of quantum superpositions,
the measurement problem and quantum advantage with quantum light. We
have discussed the averages over independent repetitions of the
experiment on a symmetrically-broken realizations, and while it seems
clear that this is the case for thermal states or RPCS which can be
built like this on purpose, one cannot exclude such superpositions to
be realized as part of one instance of the state itself, which is
indeed what happens with quantum states with no classical
counterparts. A difference would emerge between ensemble average over
independent realizations, and quantum averages over superpositions,
when bringing in interferences or information processing. For
instance, illuminating a Spatial Photonic Ising
Machines~\cite{pierangeli19a} with RPCS (which is easy) or with
multiphoton Fock states (which is hard) should lead to completely
different paradigms of computations for such devices, despite the
converging correlations for high-photon numbers between those two
cases in absence of interferences or spatial light-modulation. To
which extent thermal or random-phase coherent states still provide
additional resources for this technology is an open question to which
our approach holds the key.

Second our discussion provides a fresh perspective to the emerging
questions related to quantum coherence as a resource
theory~\cite{streltsov17a}. Recent works suggest that coherent states
with a well defined phase are useful in a quantum
context~\cite{luders21a,brune25a,brune26a}. We here provide the
perpendicular reading that such states are precisely those which are
the least connected to anything quantum, behaving exactly as
classical, independent, uncorrelated particles would. In contrast,
randomizing the phase, i.e., spoiling quantum coherence and thus
withdrawing their quantum character, brings to the fore correlations
surprisingly close to the quantum-correlated Fock states, the closer
the higher the number of photons. Our discussion therefore provides
new elements in the discussion whether a well-defined phase and purity
of the state are quantum resources or a way out of it. This is also
relevant in connection to whether they exist at all~\cite{molmer97a}.

Third, this reinforces and clarifies the central role of the coherent
state at the quantum/classical interface, a point that was already
highlighted in the Glauber/Sudarshan debates, but which did not get
fully resolved. Our description calls for deeper introspections of
interpretation of quantum mechanics along those lines. We focused on
the~$P$ function and drew a line at classically from its possible
interpretation as a probability distribution, but later works by
Drummond and Gardiner~\cite{drummond80b} introduced the positive~$P$
representation, which strips the~$P$ functions from pathologies by
dedoubling the space for the object: one can restore independence and
well-behaved probabilities but involving hidden variables in parallel
spaces. Those are familiar notions of quantum
ontology~\cite{arXiv_mckeever26a}, which may find in spatial
multiboson correlations fertile grounds for concrete, intuitive and
experimentally testable interpretations.

Last, but not least, it finally brings a resolution to the historical
conundrum that fueled so many controversies: that ``photons at two
separated detectors can ever be correlated'' is not, after all,
heretical or patently absurd. It follows, instead, from an old and
well-known statistical paradox if we accept, or even expect, that a
much more intuitive process---namely, symmetry breaking---is part of an
otherwise completely classical phenomenon.

Acknowledgements: We thank Eduardo Zubizarreta Casalengua, Joaquin
Guimbao Gaspar and Francesca Marchetti for discussions, and Carlos
Anton Solanas for putting the authors in touch with each
other. F.P.L. thanks the Cátedra Nacional de Tecnologías Cuánticas of
Colombia for their invitation to reflect on such questions, which
brought an unexpected source of inspiration. 
Support from the ``Spatial Quantum Optical Annealer for Spin
Hamiltonians'' EU Research and Innovation Project \textsc{Heisingberg}
under Grant Agreement No. 101114978 is also acknowledged.

\bibliographystyle{naturemag}
\bibliography{sci,Books,arXiv,bosoc}

\clearpage
\onecolumngrid

\setcounter{equation}{0}
\setcounter{figure}{0}
\setcounter{table}{0}
\setcounter{page}{1}
\setcounter{section}{0}

\setcounter{NAT@ctr}{0}
\renewcommand{\theequation}{S\arabic{equation}}
\renewcommand{\thefigure}{S\arabic{figure}}
\renewcommand{\theenumiv}{S\arabic{enumiv}}
\renewcommand{\bibnumfmt}[1]{[S#1]}
\renewcommand{\citenumfont}[1]{S#1}

\renewcommand{\thepage}{S\arabic{page}}

\renewcommand{\theHequation}{S\arabic{equation}}
\renewcommand{\theHfigure}{S\arabic{figure}}

\begin{center}
{\large \bfseries Boson correlations are spurious for classical states\\[.1cm]
Supplementary Material}\\[.4cm]

Daniel E.~Salazar$^{1}$ and Fabrice P.~Laussy$^{1,*}$\\[0.1cm]

{\itshape 
  ${}^{1}$Instituto de Ciencia de Materiales de Madrid (ICMM-CSIC), 28049 Madrid, Spain}\\
  ${}^{*}${fabrice.laussy@gmail.com}
\\[0.1cm]

(Dated: \today)\\[1.2cm]

{\bfseries Abstract}\\[0.4cm]
\begin{minipage}{0.85\textwidth}   
  This is the Supplementary Material for the text ``Bosons
  correlations are spurious for classical states, where we show that
  quantum states with a positive-definite~$P$ representations can be
  understood as devoid entirely of the bosonic correlations which they
  are attributed, which we explain as statistical artifacts. In a
  first part (I.~Formalism), we provide theoretical details, in
  particular keeping the expressions general in full space as opposed
  to focused on angles only in the main text. We discuss how our focus
  on single-shot realizations bring forward considerations of how
  observables have to be sampled and normalized. In a second part
  (II.~Monte Carlo illustrations), we multiply examples of all sorts
  so as to show how our main statement manifests itself in different
  configurations. Finally, in a third part (III.~Many particles), we
  similarly illustrate the mechanism with more than two particles,
  covering both as little more than that, and the case where enough
  particles are accessible to enter a new regime altogether.
\end{minipage}
\end{center}

\section{Formalism}
\subsection{Many-body wavefunctions}

Using the formalism of Zubizarreta Casalengua \&
Laussy~\citesupp{arXivSupp_casalengua24a}, the $N$-particle reduced
density matrix can be extended for~$N$ particles as:
\begin{equation}
\label{eq01}
\rho^{(N)}(\mathbf r_1,\dots,\mathbf r_N) =\bigl\langle:\hat n^{(1)}(\mathbf r_1)\cdots \hat n^{(1)}(\mathbf r_N):\bigr\rangle,
\end{equation}
where \(\hat n^{(1)}(\mathbf r)=\hat\Psi^\dagger(\mathbf r)\hat\Psi(\mathbf r)\). Expanding the field operator as \(\hat\Psi(\mathbf r)=\sum_m \phi_m(\mathbf r)\hat a_m\) and \(\hat\Psi^\dagger(\mathbf r)=\sum_m \phi_m^*(\mathbf r)\hat a_m^\dagger\), one obtains
\begin{equation}
\label{eq02}
\rho^{(N)}(\mathbf r_1,\dots,\mathbf r_N) = \sum_{\{p_j\},\{q_j\}} \left\langle \hat a_{p_1}^\dagger\cdots \hat a_{p_N}^\dagger \hat a_{q_N}\cdots \hat a_{q_1} \right\rangle \prod_{j=1}^N \phi_{p_j}^*(\mathbf r_j)\phi_{q_j}(\mathbf r_j).
\end{equation}

For two vortices rotating in opposite directions, we use the two
Laguerre--Gaussian modes
$\phi_a(\mathbf r)=\psi_0^{+\ell}(r,\theta)=R_\ell(r)e^{i\ell\theta}$
and
$\phi_b(\mathbf r)=\psi_0^{-\ell}(r,\theta)=R_\ell(r)e^{-i\ell\theta}$
with radial part
$ R_\ell(r)=1/(\sqrt{\pi |\ell|!})\,r^{|\ell|}e^{-r^2/2}$. The field
operator then reads
$ \hat\Psi(r,\theta)=R_\ell(r)\left(e^{i\ell\theta}\hat
  a+e^{-i\ell\theta}\hat b\right)$.  We further define the sign string
\begin{equation*}
S_k^N\equiv\{\underbrace{+,\dots,+}_{k},\underbrace{-,\dots,-}_{N-k}\},
\end{equation*}
and denote by \(\mathfrak S[S_k^N]\) the set of its distinct
permutations. For any \(\sigma\in\mathfrak S[S_k^N]\), \(\sigma(j)\)
denotes the sign at position \(j\). For states diagonal in the Fock
basis,
$ \hat\rho=\sum_{n_a,n_b}p_{n_a,n_b}\,|n_a,n_b\rangle\langle
n_a,n_b|$, the only nonvanishing correlators are those with equal
numbers of creation and annihilation operators in both the mode \(a\)
and mode~$b$. The corresponding modal correlators are therefore
\begin{equation}
\label{eq03}
C_{k,N-k} \equiv \left\langle (\hat a^\dagger)^k(\hat b^\dagger)^{N-k}\hat b^{\,N-k}\hat a^k
\right\rangle.
\end{equation}

The $N$-particle reduced density matrix then reads
\begin{equation}
\label{eq05}
\rho^{(N)}(\mathbf r_1,\dots,\mathbf r_N) = \frac{e^{-\sum\limits_{j=1}^N r_j^2}\prod\limits_{j=1}^N r_j^{2|\ell|}}{\pi^N(|\ell|!)^N} \sum_{k=0}^N C_{k,N-k} \sum_{\sigma_1,\sigma_2\in\mathfrak S[S_k^N]} \cos\!\left[\ell\sum_{i=1}^N\sigma_1(i)\theta_i - \ell\sum_{j=1}^N\sigma_2(j)\theta_j\right].
\end{equation}

The reduced density matrix is not normalized. Its integral over all coordinates defines
\begin{equation}
\label{eq06}
\mathcal Z_N \equiv \int d^2\mathbf r_1\cdots d^2\mathbf r_N\, \rho^{(N)}(\mathbf r_1,\dots,\mathbf r_N) = \langle\hat N(\hat N-1)\cdots(\hat N-N+1)\rangle.
\end{equation}

Since \(\phi_a(\boldsymbol{r})\) and \(\phi_b(\boldsymbol{r})\) modes are normalized, the radial sector integrates to unity, while angular integration eliminates all terms except those with \(\sigma_1=\sigma_2\). For fixed \(k\), the number of such terms is \(\binom{N}{k}\), and therefore
\begin{equation}
\label{eq07}
\mathcal Z_N = \sum_{k=0}^N \binom{N}{k}C_{k,N-k}.
\end{equation}

The normalized $N$-particle reduced density matrix, denoted with~{\raisebox{-1.5ex}{\Large\textasciitilde}}, is thus
\begin{equation}
\label{eq08}
\widetilde{\rho}^{(N)}(\mathbf r_1,\dots,\mathbf r_N) ={e^{-\sum\limits_{j=1}^N r_j^2}\prod\limits_{j=1}^N 2r_j^{2|\ell|}\over(|\ell|!)^N}\frac{\displaystyle\sum_{k=0}^N C_{k,N-k} \displaystyle\sum_{\sigma_1,\sigma_2\in\mathfrak S[S_k^N]} \cos\!\left[
\ell\sum\limits_{i=1}^N\sigma_1(i)\theta_i - \ell\sum\limits_{j=1}^N\sigma_2(j)\theta_j\right]}{(2\pi)^N\displaystyle\sum_{k=0}^N \binom{N}{k}C_{k,N-k}}.
\end{equation}

The angular dependence embeds the correlations, while the radial part
factorizes:
\begin{equation}
  \label{eq09}
  \widetilde{\rho}^{(N)}(\mathbf r_1,\dots,\mathbf r_N) =\left(\prod_{j=1}^N \vert R_\ell(r_j)\vert^2\right)\widetilde{\Theta}^{(N)}(\theta_1,\dots,\theta_N),
\end{equation}
with~$\widetilde{\Theta}^{(N)}$ providing
Eq.~(\ref{eq:SatApr4112658AMCEST2026}) of the main text.

\subsection{Many-body reduced density matrices in the \(P\) representation}\label{secIB}

A more general formulation is obtained from the Glauber--Sudarshan \(P\) representation. For a quantum state \(\rho\) expressed in a finite \(M\)-mode basis \(\{\phi_j(\mathbf r)\}_{j=1}^M\), one may write
\begin{equation}
\label{eq11}
\rho = \int d\boldsymbol{\alpha}\, P(\boldsymbol{\alpha})\, |\boldsymbol{\alpha}\rangle\langle\boldsymbol{\alpha}|,
\end{equation}
where \(d\boldsymbol{\alpha}\equiv\prod_{j=1}^M d^2\alpha_j\), \(P(\boldsymbol{\alpha})\) is the multimode Glauber--Sudarshan distribution, and $|\boldsymbol{\alpha}\rangle\langle\boldsymbol{\alpha}|=\bigotimes_{j=1}^M |\alpha_j\rangle\langle\alpha_j|$ is the tensor product of coherent states. The coherent amplitudes are written as \(\alpha_j=|\alpha_j|e^{i\varphi_j}\). The $N$-particle density matrix is
\begin{equation}
\label{eq12}
\rho^{(N)}(\mathbf r_1,\dots,\mathbf r_N) = \sum_{i_1,\dots,i_N=1}^M \sum_{j_1,\dots,j_N=1}^M \langle a_{i_1}^\dagger\cdots a_{i_N}^\dagger a_{j_N}\cdots a_{j_1} \rangle \phi_{i_1}^*(\mathbf r_1)\cdots \phi_{i_N}^*(\mathbf r_N) \phi_{j_N}(\mathbf r_N)\cdots \phi_{j_1}(\mathbf r_1).
\end{equation}
Within the \(P\) representation, the normally ordered correlators are given by the corresponding moments,
\begin{equation}
\label{eq13}
\langle a_{i_1}^\dagger\cdots a_{i_N}^\dagger a_{j_N}\cdots a_{j_1} \rangle =
\int d\boldsymbol{\alpha}\, \alpha_{i_1}^*\cdots \alpha_{i_N}^* \alpha_{j_N}\cdots \alpha_{j_1}\,
P(\boldsymbol{\alpha}).
\end{equation}
Substituting Eq.~\eqref{eq13} into Eq.~\eqref{eq12} yields
\begin{equation}
\label{eq14}
\rho^{(N)}(\mathbf r_1,\dots,\mathbf r_N) = \int d\boldsymbol{\alpha}\,P(\boldsymbol{\alpha})
\sum_{i_1,\dots,i_N=1}^M \sum_{j_1,\dots,j_N=1}^M [\alpha_{i_1}\phi_{i_1}(\mathbf r_1)]^*
\cdots [\alpha_{i_N}\phi_{i_N}(\mathbf r_N)]^* [\alpha_{j_N}\phi_{j_N}(\mathbf r_N)]
\cdots [\alpha_{j_1}\phi_{j_1}(\mathbf r_1)].
\end{equation}
Since the sums over the mode indices are independent, the expression factorizes as
\begin{equation}
\label{eq15}
\rho^{(N)}(\mathbf r_1,\dots,\mathbf r_N) = \int d\boldsymbol{\alpha}\,P(\boldsymbol{\alpha})
\prod_{i=1}^N \left|\sum_{m=1}^M \alpha_m \phi_m(\mathbf r_i) \right|^2.
\end{equation}

Each factor in the integrand is the one-particle density matrix associated with the coherent state \(|\boldsymbol{\alpha}\rangle\),
\begin{equation}
\label{eq16}
\rho_{\boldsymbol{\alpha}}^{(1)}(\mathbf r) = \left|\sum_{m=1}^M \alpha_m \phi_m(\mathbf r) \right|^2,
\end{equation}
and therefore
\begin{equation}
\label{eq17}
\rho^{(N)}(\mathbf r_1,\dots,\mathbf r_N) = \int d\boldsymbol{\alpha}\,P(\boldsymbol{\alpha})
\prod_{i=1}^N \rho_{\boldsymbol{\alpha}}^{(1)}(\mathbf r_i).
\end{equation}

For the two-mode case \((M=2)\), Eq.~\eqref{eq16} satisfies
\begin{equation}
\label{eq18}
\int d^2\mathbf r\,\rho_{\boldsymbol{\alpha}}^{(1)}(\mathbf r) = |\alpha_1|^2+|\alpha_2|^2 \equiv
n_{\boldsymbol{\alpha}}.
\end{equation}

We therefore define the normalized one-particle density matrix
\[
\widetilde{\rho}_{\boldsymbol{\alpha}}^{(1)}(\mathbf r)
\equiv
\frac{\rho_{\boldsymbol{\alpha}}^{(1)}(\mathbf r)}{n_{\boldsymbol{\alpha}}}.
\]

The normalization factor \(\mathcal Z_N\) then becomes
\begin{align}
\label{eq19}
\mathcal Z_N
&= \int d^2\mathbf r_1\cdots d^2\mathbf r_N\, \rho^{(N)}(\mathbf r_1,\dots,\mathbf r_N)\nonumber\\
&= \int d\boldsymbol{\alpha}\,P(\boldsymbol{\alpha}) \int d^2\mathbf r_1\cdots d^2\mathbf r_N \prod_{i=1}^N \rho_{\boldsymbol{\alpha}}^{(1)}(\mathbf r_i)\nonumber\\
&= \int d\boldsymbol{\alpha}\,P(\boldsymbol{\alpha}) \prod_{i=1}^N \left(\int d^2\mathbf r_i\,
\rho_{\boldsymbol{\alpha}}^{(1)}(\mathbf r_i)\right)\nonumber\\
&= \int d\boldsymbol{\alpha}\,P(\boldsymbol{\alpha})\,n_{\boldsymbol{\alpha}}^N.
\end{align}

Accordingly, the normalized $N$-particle density matrix reads
\begin{equation}
\label{eq20}
\widetilde{\rho}^{(N)}(\mathbf r_1,\dots,\mathbf r_N) = \frac{\int d\boldsymbol{\alpha}\,P(\boldsymbol{\alpha})\,n_{\boldsymbol{\alpha}}^N
\prod_{i=1}^N \widetilde{\rho}_{\boldsymbol{\alpha}}^{(1)}(\mathbf r_i)}{\int d\boldsymbol{\alpha}\,P(\boldsymbol{\alpha})\,n_{\boldsymbol{\alpha}}^N}.
\end{equation}

Throughout, \(\widetilde{\rho}\) and \(\widetilde{\Theta}\) denote
normalized reduced density matrices.

\subsection{Vortex basis}\label{sec:vortex_explicit}

For the two vortex \(\phi_a(r,\theta)=R_\ell(r)e^{i\ell\theta}\) and \(\phi_b(r,\theta)=R_\ell(r)e^{-i\ell\theta}\), the one-particle density matrix associated with a fixed coherent-state amplitude
\(\boldsymbol{\alpha}=(\alpha_1,\alpha_2)^T\) is
\begin{equation}
\label{eq21}
\rho_{\boldsymbol{\alpha}}^{(1)}(r,\theta) = |R_\ell(r)|^2\, \bigl|\alpha_1 e^{i\ell\theta}+\alpha_2 e^{-i\ell\theta}\bigr|^2.
\end{equation}
and therefore, the normalized density matrix is
\begin{equation}
\label{eq22}
\widetilde{\rho}_{\boldsymbol{\alpha}}^{(1)}(r,\theta) = |R_\ell(r)|^2\, \frac{\bigl|
\alpha_1 e^{i\ell\theta}+\alpha_2 e^{-i\ell\theta}\bigr|^2}{|\alpha_1|^2+|\alpha_2|^2}.
\end{equation}

It is convenient to introduce the variables
\(\alpha_1\equiv\sqrt{st}\,e^{i\varphi_1-\eta}\), and
\( \alpha_2\equiv\sqrt{s(1-t)}\,e^{i\varphi_2-\eta}\) with
\(s\in[0,\infty),~t\in[0,1]\), and
\(\eta\equiv\varphi_1-\varphi_2\). In these variables,
\(n_{\boldsymbol{\alpha}}=s\) given by Eq.~(\ref{eq18}) and
\begin{equation*}
\frac{2|\alpha_1||\alpha_2|}{|\alpha_1|^2+|\alpha_2|^2} = 2\sqrt{t(1-t)}.
\end{equation*}
Equation~\eqref{eq22} then becomes
\begin{equation}
\label{eq23}
\widetilde{\rho}_{t}^{(1)}(r,\theta) = |R_\ell(r)|^2 \Bigl[1+2\sqrt{t(1-t)}\cos\!\bigl(2\ell\theta\bigr) \Bigr].
\end{equation}

A key simplification is that \(\widetilde{\rho}_{t}^{(1)}\) is independent of
the total intensity variable \(s\). The integration transforms as \(d^2\alpha_1\,d^2\alpha_2= 1/4\,s\,ds\,dt\,d\varphi_1\,d\varphi_2\).
Substituting Eq.~\eqref{eq23} into Eq.~\eqref{eq20}, the normalized \(N\)-particle
reduced density matrix may be written as
\begin{equation}
\label{eq24}
\widetilde{\rho}^{(N)}(\mathbf r_1,\dots,\mathbf r_N) = \frac{\displaystyle \int_0^\infty ds\int_0^1 dt\int_0^{2\pi} d\varphi_1\int_0^{2\pi} d\varphi_2\,\frac{s}{4}\, P\!\left(\sqrt{st}\,e^{i\varphi_1},\sqrt{s(1-t)}\,e^{i\varphi_2}\right) s^{N} \prod_{j=1}^{N}
\widetilde{\rho}_{t}^{(1)}(r_j,\theta_j+\eta)}{\displaystyle \int_0^\infty ds\int_0^1 dt\int_0^{2\pi} d\varphi_1\int_0^{2\pi} d\varphi_2\, \frac{s}{4}\,P\!\left(\sqrt{st}\,e^{i\varphi_1},\sqrt{s(1-t)}\,e^{i\varphi_2}\right)s^{N}}.
\end{equation}

For states that do not fix a global phase, the only relevant phase is
the relative phase $\eta$. For instance, in the balanced (i.e., same
population) thermal case one has
\begin{equation}
\label{eq25}
P_{\rm th}(\alpha_1,\alpha_2) =\frac{1}{(\pi\bar n)^2} \exp\!\left[-\frac{|\alpha_1|^2+|\alpha_2|^2}{\bar n}\right],
\end{equation}
so that, after integration over the global phase, the remaining integrals involve the factor \(s^{N+1}/\bar n^2 e^{-s/\bar n} ds\,dt\,d\eta/2\pi\) and the $s$-integral cancels between numerator and denominator, yielding
\begin{equation}
\label{eq26}
\widetilde{\rho}^{(N)}_{\rm th}(\mathbf r_1,\dots,\mathbf r_N) = \int_0^1 dt \int_0^{2\pi}\frac{d\eta}{2\pi} \prod_{j=1}^{N}\widetilde{\rho}_{t}^{(1)}(r_j,\theta_j+\eta)
\end{equation}
as given in the main text.  In the imbalanced thermal case, by
contrast, the \(s\)-integration produces an \(N\)-dependent effective
weight

\begin{equation}
\label{eq27}
W_N(t)=\frac{A(t)^{-(N+2)}}{\displaystyle \int_0^1 dt\,A(t)^{-(N+2)}}\,,
\end{equation}
where
\begin{equation}
  \label{eq:FriApr17030345PMCEST2026}
A(t)\equiv\frac{t}{\bar n_1}+\frac{1-t}{\bar n_2}\,.  
\end{equation}
%
so that
\begin{equation}
\label{eq28}
\widetilde{\rho}^{(N)}_{\rm th}(\mathbf r_1,\dots,\mathbf r_N) = \int_0^1 dt\,W_N(t) \int_0^{2\pi}\frac{d\eta}{2\pi} \prod_{j=1}^{N} \widetilde{\rho}_{t}^{(1)}(r_j,\theta_j+\eta).
\end{equation}

Thus, in the vortex basis the normalized \(N\)-particle reduced
density matrix can be interpreted as an average over shared geometries
\((t,\eta)\), with the role of \(s\) reduced to an effective
statistical weight. The physical conclusions are discussed in the main
text.

\subsection{Dipolar basis}\label{sec:dipolar}

The above results are general and can be expressed in any basis, for
any number of modes and any number of photons. Here we illustrate the
change of basis by working in the dipolar (Cartesian) basis. In this
case, one also needs to retain the full spatial correlations and to
sample jointly the radius and angle of each particle, while still
sampling the various particles independently the ones from the
others.  
We now specialize the general multimode expressions,
Eqs.~\eqref{eq01}--\eqref{eq02}, to the first-order Hermite--Gauss
subspace spanned by the dipole modes
\begin{equation}
u_x(x,y)=\mathcal N\,x\,e^{-(x^2+y^2)/2},
\qquad
u_y(x,y)=\mathcal N\,y\,e^{-(x^2+y^2)/2},
\label{eq29}
\end{equation}
where \(\mathcal N\) is fixed by orthonormality and
\(\mathbf r\equiv(x,y)\). The field operator reads
\(\hat\Psi(\mathbf r)=u_x(\mathbf r)\hat a_x+u_y(\mathbf r)\hat
a_y\). For states diagonal in the dipolar Fock basis, i.e., with
\(\hat\rho=\sum_{n_x,n_y}p_{n_x,n_y}\,|n_x,n_y\rangle\langle
n_x,n_y|\), the only nonvanishing correlators are those with equal
numbers of creation and annihilation operators in mode \(x\). We
define the dipolar string
\begin{equation*}
T_k^N\equiv\{\underbrace{x,\dots,x}_{k},\underbrace{y,\dots,y}_{N-k}\},
\end{equation*}
and denote by \(\mathfrak S[T_k^N]\) the set of its distinct permutations. For any \(\mu\in\mathfrak S[T_k^N]\), \(\mu(j)\) denotes the mode label at position \(j\). The corresponding modal correlators are therefore
\begin{equation}
\label{eq30}
C_{k,N-k} = \left\langle (\hat a_x^\dagger)^k(\hat a_y^\dagger)^{N-k}\hat a_y^{\,N-k}\hat a_x^k \right\rangle.
\end{equation}

Since the dipole modes may be chosen real, the \(N\)-particle reduced density matrix takes the form
\begin{equation}
\label{eq31}
\rho^{(N)}(\mathbf r_1,\dots,\mathbf r_N) = \sum_{k=0}^N C_{k,N-k}
\sum_{\mu_1,\mu_2\in\mathfrak S[T_k^N]} \prod_{j=1}^N u_{\mu_1(j)}(\mathbf r_j)\,u_{\mu_2(j)}(\mathbf r_j).
\end{equation}
This is the exact analogue of Eq.~\eqref{eq05} for the vortex basis,
except that the spatial correlations are no longer purely angular. The
reduced density matrix is again not normalized. Using the
orthonormality of the single-particle modes, we find again
\begin{equation}
\label{eq32}
\mathcal Z_N = \int d^2\mathbf r_1\cdots d^2\mathbf r_N\,\rho^{(N)}(\mathbf r_1,\dots,\mathbf r_N) = \sum_{k=0}^N \binom{N}{k} C_{k,N-k},
\end{equation}
since all terms vanish upon integration except those with \(\mu_1=\mu_2\). The normalized \(N\)-particle reduced density matrix is therefore
\begin{equation}
\label{eq33}
\widetilde\rho^{(N)}(\mathbf r_1,\dots,\mathbf r_N) = \frac{\displaystyle\sum_{k=0}^N C_{k,N-k}\sum_{\mu_1,\mu_2\in\mathfrak S[T_k^N]} \prod_{j=1}^N u_{\mu_1(j)}(\mathbf r_j)\,u_{\mu_2(j)}(\mathbf r_j)}{\displaystyle\sum_{k=0}^N \binom{N}{k} C_{k,N-k}}.
\end{equation}

The same object can be recast in the Glauber--Sudarshan \(P\)
representation. For a fixed coherent-state realization
\(\boldsymbol\beta=(\beta_x,\beta_y)^T\), the associated one-particle
density matrix reads
\(\rho_\beta^{(1)}(x,y)=|\beta_x u_x(x,y)+\beta_y u_y(x,y)|^2\), that
is,
\begin{equation}
\rho_\beta^{(1)}(x,y) = \mathcal N^2 e^{-(x^2+y^2)} \left(|\beta_x|^2 x^2 + |\beta_y|^2 y^2 + 2\,\mathrm{Re}\!\left(\beta_x\beta_y^*\right)xy \right).
\label{eq34}
\end{equation}
By orthonormality,
\begin{equation}
n_\beta = \int dx\,dy\,\rho_\beta^{(1)}(x,y) = |\beta_x|^2+|\beta_y|^2,
\label{eq35}
\end{equation}
and the normalized one-particle density matrix is
\begin{equation}
\widetilde\rho_\beta^{(1)}(x,y) = \frac{\rho_\beta^{(1)}(x,y)}{n_\beta}.
\label{eq36}
\end{equation}

Equation~\eqref{eq34} shows that a generic realization is not rotationally symmetric, since it is governed by an anisotropic quadratic form in \((x,y)\). A single realization becomes rotationally symmetric only when \(|\beta_x|=|\beta_y|\) and \(\mathrm{Re}(\beta_x\beta_y^*)=0\), namely for relative phase \(\pm\pi/2\), i.e., \(\beta_y=\pm i\,\beta_x\). Only in this case does the one-particle density matrix reduce to a donut-like profile proportional to \(e^{-(x^2+y^2)}(x^2+y^2)\). In complete analogy with the vortex basis, we introduce the variables \(\beta_x\equiv\sqrt{st}\,e^{i\varphi_x}\) and \(\beta_y\equiv\sqrt{s(1-t)}\,e^{i\varphi_y}\), with \(s\in[0,\infty)\), \(t\in[0,1]\), and \(\eta\equiv\varphi_x-\varphi_y\). Then \(n_\beta=s\), while \( \mathrm{Re}(\beta_x\beta_y^*)=s\sqrt{t(1-t)}\cos\eta\),
so that Eq.~\eqref{eq36} becomes
\begin{equation}
\label{eq37}
\widetilde\rho_{t,\eta}^{(1)}(x,y) = \mathcal N^2 e^{-(x^2+y^2)} \left[t\,x^2+(1-t)\,y^2+2\sqrt{t(1-t)}\cos\eta\,xy \right].
\end{equation}
In particular, \(\widetilde\rho_{t,\eta}^{(1)}\) is independent of the total intensity \(s\). Using \(d^2\beta_x\,d^2\beta_y=1/4\,s\,ds\,dt\,d\varphi_x\,d\varphi_y\), the normalized \(N\)-particle reduced density matrix becomes
\begin{equation}
\label{eq38}
\widetilde\rho^{(N)}(\mathbf r_1,\dots,\mathbf r_N) = \frac{\displaystyle \int_0^\infty ds\int_0^1 dt\int_0^{2\pi}d\varphi_x\int_0^{2\pi}d\varphi_y\,\frac{s}{4}\, P\!\left(\sqrt{st}\,e^{i\varphi_x},\sqrt{s(1-t)}\,e^{i\varphi_y}\right)s^N\prod_{j=1}^N \widetilde\rho_{t,\eta}^{(1)}(x_j,y_j)}{\displaystyle\int_0^\infty ds\int_0^1 dt\int_0^{2\pi}d\varphi_x\int_0^{2\pi}d\varphi_y\,\frac{s}{4}\,P\!\left(\sqrt{st}\,e^{i\varphi_x},\sqrt{s(1-t)}\,e^{i\varphi_y}\right)s^N}.
\end{equation}

For states that do not fix a global phase, the two phase integrals reduce to an average over the relative phase \(\eta\). The resulting structure is therefore identical to that obtained in the vortex basis, the normalized \(N\)-particle reduced density matrix is an average over shared geometries \((t,\eta)\), while the total intensity \(s\) enters only through the effective \(N\)-particle statistical weight. Consequently, the balanced and imbalanced thermal cases follow exactly as in Sec.~\ref{sec:vortex_explicit}, the only difference being the explicit form of the conditioned one-particle density matrix in Eq.~\eqref{eq37}.

\subsection{Normalization and sampling}

Analytical expectation values follow directly from the normalized
reduced density matrix in Eq.~(\ref{eq08}), or equivalently from its
angular part in Eq.~(\ref{eq:SatApr4112658AMCEST2026}) whenever the
observable depends only on the azimuthal coordinates. For a
\(q\)-particle observable \(F(\mathbf r_1,\dots,\mathbf r_q)\),
\begin{equation}
\label{eq39}
\langle F\rangle^{(q)} = \int d^2\mathbf r_1\cdots d^2\mathbf r_q\, F(\mathbf r_1,\dots,\mathbf r_q)\, \widetilde{\rho}^{(q)}(\mathbf r_1,\dots,\mathbf r_q).
\end{equation}

If \(F\) depends only on the angular variables, this reduces
to~$\langle F\rangle^{(q)} = \int_0^{2\pi} d\theta_1\cdots
\int_0^{2\pi} d\theta_q\, F(\theta_1,\dots,\theta_q)\,
\widetilde{\Theta}^{(q)}(\theta_1,\dots,\theta_q)$. For mixed states
diagonal in the Fock basis
\(\hat\rho =\sum_{n_a,n_b} p_{n_a,n_b}\,|n_a,n_b\rangle\langle
n_a,n_b|,\) with \(\sum_{n_a,n_b} p_{n_a,n_b}=1,\) the reduced density
matrix is linear in the state,
\begin{equation}
\label{eq41}
\rho^{(q)}_{\hat\rho} = \sum_{n_a,n_b} p_{n_a,n_b}\,
\rho^{(q)}_{|n_a,n_b\rangle},
\end{equation}
however, the corresponding normalized quantity is not linear, since
\begin{equation}
\label{eq42}
\widetilde{\rho}^{(q)}_{\hat\rho} = \frac{\rho^{(q)}_{\hat\rho}}{\mathcal Z_q[\hat\rho]},
\qquad
\mathcal Z_q[\hat\rho] =\int d^2\mathbf r_1\cdots d^2\mathbf r_q\,
\rho^{(q)}_{\hat\rho}(\mathbf r_1,\dots,\mathbf r_q).
\end{equation}
Therefore, statistical mixing and normalization do not commute:
\begin{equation}
\label{eq43}
\widetilde{\rho}^{(q)}_{\hat\rho}\neq\sum_{n_a,n_b} p_{n_a,n_b}\, \widetilde{\rho}^{(q)}_{|n_a,n_b\rangle}.
\end{equation}

To make this explicit, let $N=n_a+n_b$ denote the total occupation of a given Fock sector. For the state $|n_a,n_b\rangle$, Eq.~(\ref{eq06}) yields \(\mathcal Z_q[n_a,n_b] = N!/(N-q)!\) with \(N\ge q\), and $\mathcal Z_q[n_a,n_b]=0$ otherwise. Hence
\begin{equation}
\label{eq44}
\widetilde{\rho}^{(q)}_{\hat\rho} = \frac{\displaystyle\sum_{n_a,n_b} p_{n_a,n_b}\,
\mathcal Z_q[n_a,n_b]\,\widetilde{\rho}^{(q)}_{|n_a,n_b\rangle}}{\displaystyle\sum_{n_a,n_b} p_{n_a,n_b}\,\mathcal Z_q[n_a,n_b]}.
\end{equation}

An identical relation holds for the angular distribution,
\begin{equation}
\label{eq45}
\widetilde{\Theta}^{(q)}_{\hat\rho} = \frac{\displaystyle\sum_{n_a,n_b} p_{n_a,n_b}\,
\mathcal Z_q[n_a,n_b]\, \widetilde{\Theta}^{(q)}_{|n_a,n_b\rangle}}{\displaystyle\sum_{n_a,n_b} p_{n_a,n_b}\,\mathcal Z_q[n_a,n_b]}.
\end{equation}

Equations~(\ref{eq44}) and~(\ref{eq45}) show that the normalized $q$-body distribution of a mixture is a conditional distribution: it is weighted not only by the preparation probabilities $p_{n_a,n_b}$, but also by the total number of ordered $q$-tuples contributed by each sector. In other words, the relevant effective weights are
\begin{equation}
\label{eq46}
p^{(q)}_{n_a,n_b} = \frac{p_{n_a,n_b}\,\mathcal Z_q[n_a,n_b]}{\sum_{n_a',n_b'}p_{n_a',n_b'}\,\mathcal Z_q[n_a',n_b']},
\end{equation}
so that
\begin{equation}
\label{eq47}
\widetilde{\Theta}^{(q)}_{\hat\rho} = \sum_{n_a,n_b} p^{(q)}_{n_a,n_b}\,\widetilde{\Theta}^{(q)}_{|n_a,n_b\rangle}.
\end{equation}
This point is essential for thermal states and, more generally, for any state with number fluctuations. In the Glauber--Sudarshan \(P\) representation, the same conditional normalization appears in continuous form, leading naturally to an effective \(q\)-body measure over the shared geometry \(\boldsymbol{\alpha}\). Taking Eq.~\eqref{eq20} to order \(q\), one obtains
\begin{equation}
\label{eq48}
\widetilde{\rho}^{(q)}(\mathbf r_1,\dots,\mathbf r_q) = \frac{\int d\boldsymbol{\alpha}\, P(\boldsymbol{\alpha})\,n_{\boldsymbol{\alpha}}^{q}\displaystyle\prod_{i=1}^{q} \widetilde{\rho}_{\boldsymbol{\alpha}}^{(1)}(\mathbf r_i)}{\int d\boldsymbol{\alpha}\, P(\boldsymbol{\alpha})\,n_{\boldsymbol{\alpha}}^{q}},
\end{equation}
with $n_{\boldsymbol{\alpha}}$ given by Eq.~\eqref{eq18}. It is therefore natural to define the effective \(q\)-body measure
\begin{equation}
\label{eq49}
Q_q(\boldsymbol{\alpha})\equiv\frac{P(\boldsymbol{\alpha})\,n_{\boldsymbol{\alpha}}^{q}}{
\int d\boldsymbol{\alpha}'\,P(\boldsymbol{\alpha}')\,n_{\boldsymbol{\alpha}'}^{q}},
\end{equation}
so that Eq.~\eqref{eq42} can be rewritten as
\begin{equation}
\label{eq50}
\widetilde{\rho}^{(q)}(\mathbf r_1,\dots,\mathbf r_q) =\int d\boldsymbol{\alpha}\,
Q_q(\boldsymbol{\alpha})\prod_{i=1}^{q}\widetilde{\rho}_{\boldsymbol{\alpha}}^{(1)}(\mathbf r_i).
\end{equation}
Equation~\eqref{eq50} makes explicit that the normalized \(q\)-body reduced density matrix is a conditional mixture over geometries \(\boldsymbol{\alpha}\), with weights given by \(Q_q\), not by the original \(P\) distribution. This is the continuous analogue of the Fock-basis statement that the normalized \(q\)-body distribution of a mixture is weighted by the total number of ordered \(q\)-tuples contributed by each number sector. 

For any \(q\)-body observable \(F(\mathbf r_1,\dots,\mathbf r_q)\), its expectation value follows from Eq.~\eqref{eq39}
or equivalently,
\begin{equation}
\label{eq51}
\langle F\rangle^{(q)} = \int d\boldsymbol{\alpha}\, Q_q(\boldsymbol{\alpha})\,F_{\boldsymbol{\alpha}}^{(q)},
\end{equation}
where
\begin{equation}
\label{eq52}
F_{\boldsymbol{\alpha}}^{(q)} \equiv \int d^2\mathbf r_1\cdots d^2\mathbf r_q\, F(\mathbf r_1,\dots,\mathbf r_q) \prod_{i=1}^{q} \widetilde{\rho}_{\boldsymbol{\alpha}}^{(1)}(\mathbf r_i).
\end{equation}
The connection between Eqs.~(\ref{eq41})--(\ref{eq47}) and Eqs.~(\ref{eq48})--(\ref{eq51}) is immediate once one recognizes that both describe the same conditional \(q\)-body normalization in two different representations. In the Fock decomposition, a sector \(|n_a,n_b\rangle\) contributes with weight
\[
\mathcal Z_q[n_a,n_b]=\frac{(n_a+n_b)!}{(n_a+n_b-q)!},
\]
namely, with the total number of ordered \(q\)-tuples supported by that sector. In the Glauber--Sudarshan representation, the corresponding weight is \(n_{\boldsymbol{\alpha}}^q\), so that the effective measure becomes \(Q_q(\boldsymbol{\alpha})\propto P(\boldsymbol{\alpha})n_{\boldsymbol{\alpha}}^q\). 

\subsection{Monte Carlo sampling}

The same procedure governs numerical sampling. Suppose that each
realization is generated by first drawing a shared geometry
\(\boldsymbol{\alpha}_f\), and then sampling particle positions
independently from the conditional one-particle density
\(\widetilde{\rho}^{(1)}_{\boldsymbol{\alpha}_f}\). If the geometries
are drawn from the original Glauber--Sudarshan distribution
\(P(\boldsymbol{\alpha})\), the estimation of the \(q\)-body
observable~$F_P^{(q)}$ measured from the quantum state with
Glauber~$P$ distribution must reproduce the effective measure
\(Q_q(\boldsymbol{\alpha})\propto
P(\boldsymbol{\alpha})n_{\boldsymbol{\alpha}}^q\). This may be
implemented either through explicit reweighting,


\begin{equation}
\label{eq53}
{F}^{(q)}_{P} = \frac{\sum_{f=1}^{N_{\rm fr}} n_{\boldsymbol{\alpha}_f}^{q}\,A_f^{(q)}}{
\sum_{f=1}^{N_{\rm fr}} n_{\boldsymbol{\alpha}_f}^{q}},
\end{equation}
and sampling~$\boldsymbol{\alpha}_f$ from~$P$
\begin{equation}
  \label{eq:FriApr17034809PMCEST2026}
\boldsymbol{\alpha}_f\sim P\,,
\end{equation}
or, equivalently, through averaging over all detected
\(q\)-tuples. Indeed, let the \(f\)th realization contain \(M_f\)
detected particles, and let \(A_f^{(q)}\) denote the average of the
observable over the \(\binom{M_f}{q}\) unordered \(q\)-tuples
contained in that realization. Then the corresponding estimator now
reads
\begin{equation}
\label{eq54}
{F}'^{(q)}_{\rm tuples} = \frac{\sum_f \binom{M_f}{q}\,A_f^{(q)}}{\sum_f \binom{M_f}{q}}.
\end{equation}

We define a symmetric \(q\)-body observable as one that remains
invariant under any permutation of the \(q\) detected particles, i.e.,
$F(\mathbf r_1,\dots,\mathbf r_q)=F(\mathbf r_{\pi(1)},\dots,\mathbf
r_{\pi(q)})$ for every permutation \(\pi\), in which case ordered and
unordered \(q\)-tuples yield the same value of the observable.
Passing from one convention to the other only introduces an overall
factor \(q!\), which cancels between the numerator and denominator of
the estimator.  The relation between Eqs.~(\ref{eq53})
and~(\ref{eq54}) becomes explicit when the conditional multiplicity is
Poissonian. If
\(M_f|\boldsymbol{\alpha}_f\sim{\rm
  Poisson}(n_{\boldsymbol{\alpha}_f})\), then
\begin{equation}
\label{eq55}
\mathbb E\!\left[\binom{M_f}{q}\middle|\boldsymbol{\alpha}_f\right] = \frac{n_{\boldsymbol{\alpha}_f}^{q}}{q!},
\end{equation}
so that tuple averaging induces, on average, the same geometric weighting as the explicit factor \(n_{\boldsymbol{\alpha}}^q\). Alternatively, one may sample the shared geometries directly from the effective distribution \(Q_q(\boldsymbol{\alpha})\). In that case, the \(q\)-body weighting is already built into the frame ensemble, and the unbiased estimator reduces to the simple frame average
\begin{equation}
\label{eq56}
\widehat{F}^{(q)}_{Q_q} = \frac{1}{N_{\rm fr}} \sum_{f=1}^{N_{\rm fr}} A_f^{(q)},
\qquad
\boldsymbol{\alpha}_f\sim Q_q.
\end{equation}

Any additional weighting by \(n_{\boldsymbol{\alpha}}^q\) or by \(\binom{M_f}{q}\) would then amount to a double counting of the same \(q\)-body normalization. Therefore, the relevant ensemble for a \(q\)-body observable is the ensemble of detected \(q\)-tuples. Whether this weighting is implemented explicitly through \(n_{\boldsymbol{\alpha}}^q\), implicitly through \(Q_q\), or empirically through tuple counting is a matter of sampling protocol rather than principle. 

In the Fock-basis formulation, the same distinction arises without explicitly invoking the \(P\) representation. One may either sample directly from the normalized \(q\)-body reduced density matrix \(\widetilde{\rho}^{(q)}_{\hat\rho}\), or sample full realizations of the original mixed state \(\hat\rho=\sum_{n_a,n_b}p_{n_a,n_b}|n_a,n_b\rangle\langle n_a,n_b|\). In the former case, the Fock sector must be drawn with the effective \(q\)-body weights
\[
p^{(q)}_{n_a,n_b} = \frac{p_{n_a,n_b}\,\mathcal Z_q[n_a,n_b]}{\sum_{n_a',n_b'}
p_{n_a',n_b'}\,\mathcal Z_q[n_a',n_b']},
\]
so that each sampled configuration is already a realization of the correctly normalized \(q\)-body distribution and no additional weighting is required in the analysis. In the latter case, the sectors are drawn with the preparation weights \(p_{n_a,n_b}\), and a \(q\)-body observable must then be estimated by averaging over all detected \(q\)-tuples within each realization, equivalently weighting each realization by the number of \(q\)-tuples it contributes.

\subsection{Pair-distance distribution}\label{sec:distance_vortices}

\begin{figure}[htbp]
  \centering
  \includegraphics[width=\textwidth]{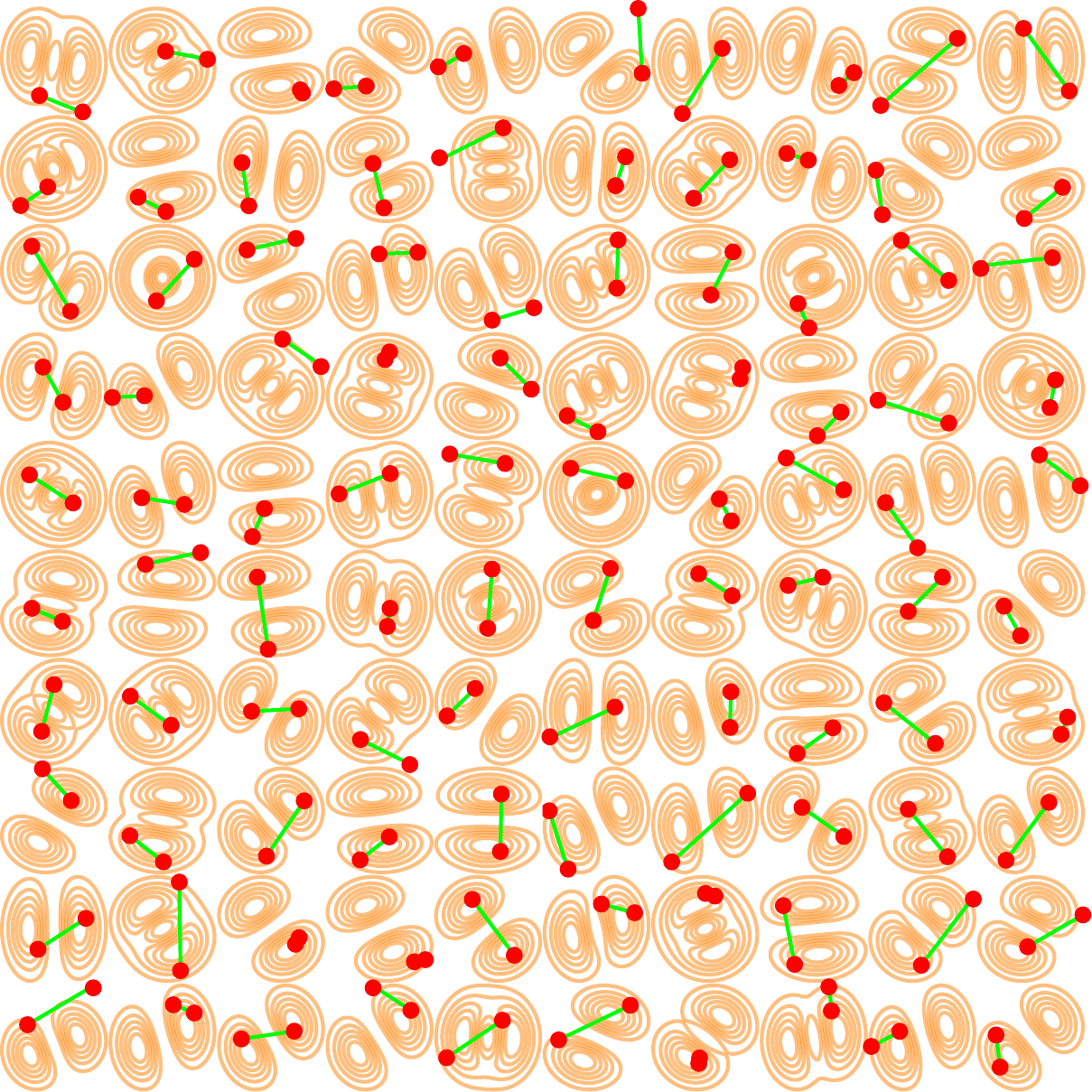}
  \caption{A \(10\times10\) grid of single-shot measurements for the
    balanced thermal state in the vortex basis \(\ell=\pm1\). Each
    panel shows a realization of the underlying geometry, which is
    sampling by Monte Carlo a distorted dipole from the thermal~$P$
    distribution, that is shown as a countour plot to capture its
    topology.  From each distribution, two points are subsequently and
    independently sampled, and marked in red, while their
    distance---the observable---is shown in green. As one can see,
    each realization is greatly fluctuating, but the two points that
    are produced are always uncorrelated. Each pair now put back in
    the total ensemble---which forms a donut---appears to be
    correlated. This is traditionally understood as boson bunching or
    boson correlations, but is merely a manifestation of spurious
    correlations from the amalgamation paradox, itself a particular
    case of the Simpson paradox.}
  \label{fignew1}
\end{figure}

A particularly insightful two-body observable for spatial correlations
is the distances between the particles detected, which follows from
the reduced two-body density matrix~$\tilde\rho^{(2)}$ as
\begin{equation}
\label{eq57}
D(d) = \int d^2\mathbf r_1\, d^2\mathbf r_2\, \widetilde\rho^{(2)}(\mathbf r_1,\mathbf r_2)\,\delta\!\left(d-|\mathbf r_1-\mathbf r_2|\right),
\end{equation}
which is normalized to unity \(\int_0^\infty dd\, D(d)=1\). Equivalently, in terms of the unnormalized reduced density matrix,
\begin{equation}
\label{eq58}
D(d) = \frac{\int d^2\mathbf r_1\, d^2\mathbf r_2\, \rho^{(2)}(\mathbf r_1,\mathbf r_2)\,
\delta\!\left(d-|\mathbf r_1-\mathbf r_2|\right)}{\int d^2\mathbf r_1\, d^2\mathbf r_2\,
\rho^{(2)}(\mathbf r_1,\mathbf r_2)}.
\end{equation}

For the counter-rotating vortex pair LG$_{p=0}^{\ell=\pm1}$, the normalized two-particle reduced density matrix factorizes into radial and angular parts as $\widetilde\rho^{(2)}(\mathbf r_1,\mathbf r_2) = \widetilde\Theta^{(2)}(\theta_1,\theta_2) \prod_{j=1}^2 |R_1(r_j)|^2$, 
with \(|R_1(r)|^2=r^2 e^{-r^2}/\pi\) and \( \widetilde\Theta^{(2)}(\theta_1,\theta_2) = 1/(2\pi)^2\Bigl[1+2\mathscr C\cos(2\Delta\theta)\Bigr]\), where \( \Delta\theta\equiv\theta_2-\theta_1\) and \(\mathscr{C}\equiv{C_{11}}\big/({C_{20}+2C_{11}+C_{02}})\). Therefore,
\begin{equation}
\label{eq60}
\widetilde\rho^{(2)}(\mathbf r_1,\mathbf r_2) = \frac{r_1^2 r_2^2 e^{-r_1^2-r_2^2}}{\pi^2}
\Bigl[1+2\mathscr C\cos(2\Delta\theta)\Bigr].
\end{equation}
Substituting Eq.~\eqref{eq60} into Eq.~\eqref{eq57}, using \(|\mathbf r_1-\mathbf r_2| = \sqrt{r_1^2+r_2^2-2r_1r_2\cos\Delta\theta}\), and exploiting rotational invariance, one obtains
\begin{equation}
\label{eq61}
D(d) = \int_0^\infty dr_1\,2r_1^3 e^{-r_1^2} \int_0^\infty dr_2\,2r_2^3 e^{-r_2^2} \int_0^\pi \frac{d\Delta\theta}{\pi} \Bigl[1+2\mathscr C\cos(2\Delta\theta)\Bigr] \delta\!\left(d-\sqrt{r_1^2+r_2^2-2r_1r_2\cos\Delta\theta}\right).
\end{equation}
Carrying out the integrations yields the closed form
\begin{equation}
\label{eq62}
D(d) = \frac{d}{16} \Bigl[(1+2\mathscr C)(8+d^4)-16\mathscr C\,d^2\Bigr]
e^{-d^2/2}.
\end{equation}
Equation~\eqref{eq62} makes explicit how the two-body distance distribution is entirely controlled by the same correlation parameter \(\mathscr C\) that appears in the angular wavefunction. In particular, for independent particles sampled from the same donut profile, \(\mathscr C=0\), and one recovers
\begin{equation}
\label{eq63}
D^\circledcirc_{\mathrm{ind}}(d) = \frac{d}{16}(8+d^4)e^{-d^2/2},
\end{equation}
whereas for the two-photon Fock state \(\ket{1\circlearrowleft,1\circlearrowright}\), for which \(\mathscr C=1/2\), one finds
\begin{equation}
\label{eq64}
D_{\ket{1,1}}(d) = \frac{d}{8}(8-4d^2+d^4)e^{-d^2/2}.
\end{equation}

The same observable admits a particularly transparent interpretation in the \(P\) representation. Defining
\begin{equation}
\label{eq65}
D_{\boldsymbol{\alpha}}(d) = \int d^2\mathbf r_1\, d^2\mathbf r_2\,\widetilde\rho_{\boldsymbol{\alpha}}^{(1)}(\mathbf r_1)\, \widetilde\rho_{\boldsymbol{\alpha}}^{(1)}(\mathbf r_2)\, \delta\!\left(d-|\mathbf r_1-\mathbf r_2|\right),
\end{equation}
the physical distance distribution may be written as
\begin{equation}
\label{eq66}
D(d) = \frac{\int d^2\alpha_1\,d^2\alpha_2\, P(\alpha_1,\alpha_2)\,n_{\boldsymbol{\alpha}}^2\,D_{\boldsymbol{\alpha}}(d)}{\int d^2\alpha_1\,d^2\alpha_2\, P(\alpha_1,\alpha_2)\,n_{\boldsymbol{\alpha}}^2}.
\end{equation}
This is the form used in the Monte Carlo sampling discussed below:
particles are drawn independently within each shared geometry
\(\boldsymbol{\alpha}\), and the resulting histogram of pair distances
reproduces the analytical \(D(d)\) obtained from the full two-particle
reduced density matrix. Figure~\ref{fignew1} illustrates explicitly
the Monte Carlo construction of distribution of distances between
pairs of points from balanced thermal states (i.e., with same
populations for both modes). In each realization, a shared geometry
\(\boldsymbol{\alpha}=(\alpha_1,\alpha_2)\) is first sampled from the
thermal \(P\) distribution, and two particle positions are then
sampled independently from the corresponding one-body density
\(\widetilde\rho^{(1)}_{\boldsymbol{\alpha}}(\mathbf r)\). Their
separation contributes one event to the histogram of pair
distances. Repeating this procedure over many realizations
reconstructs the thermal \(D(d)\) of Eq.~\eqref{eq66}. The figure
therefore makes explicit that, in the balanced thermal case, the
observed two-body correlations arise from averaging over the ensemble
of shared geometries, while the particles remain conditionally
independent within each realization. This is Simpson's paradox of
aggregates producing correlations not present in its subgroups.

A similar construction can be carried out for the dipolar
basis. Starting from the general definition given by Eq.~(\ref{eq57}),
we take the case \(N=2\) of Eq.~\eqref{eq33}. One then obtains
\begin{equation}
\label{eq67}
\widetilde\rho^{(2)}(\mathbf r_1,\mathbf r_2) = \frac{C_{20}\,u_x^2(\mathbf r_1)u_x^2(\mathbf r_2) + C_{02}\,u_y^2(\mathbf r_1)u_y^2(\mathbf r_2) + C_{11}\Big(u_x(\mathbf r_1)u_y(\mathbf r_2)+u_y(\mathbf r_1)u_x(\mathbf r_2)\Big)^2}{C_{20}+2C_{11}+C_{02}}.
\end{equation}
\begin{figure}[htbp]
  \centering
  \includegraphics[width=0.6\textwidth]{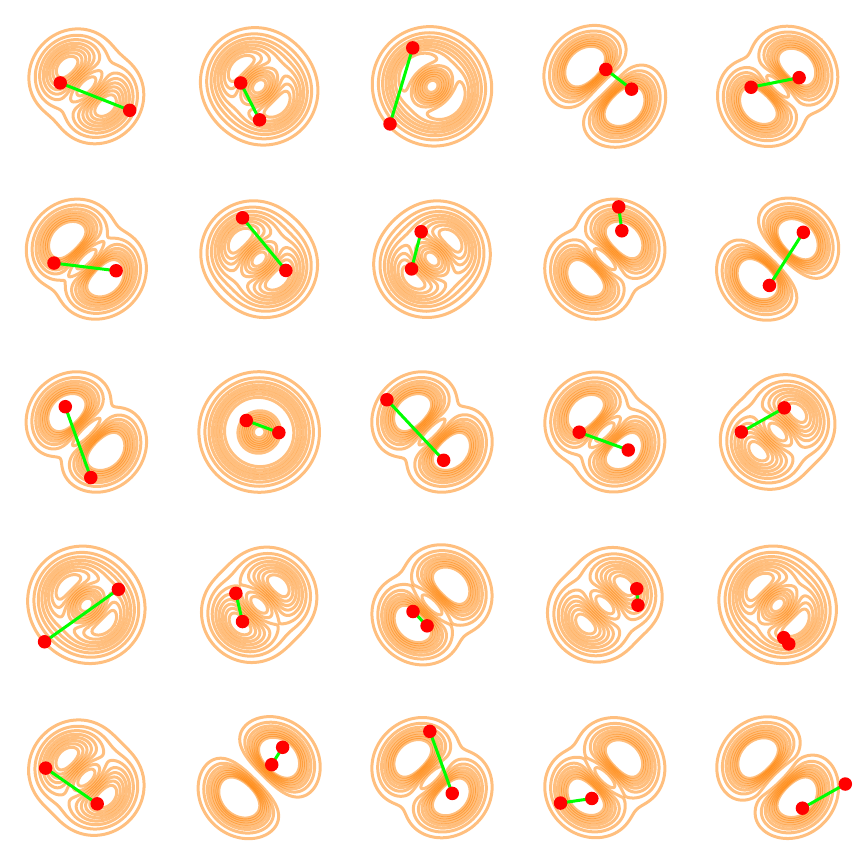}
  \caption{A \(5\times5\) grid of single-shot patterns for the RPCS
    state in the dipolar basis, i.e., interfering an horizontal dipole
    with a vertical one. Each panel shows a geometry drawn from the
    corresponding distribution \(P(\beta_x,\beta_y)\), together with
    two independently sampled positions marked in red and their mutual
    distance shown in green. The superposition of the two sampled
    points from all panels, viewed as a single aggregate frame, is
    shown in panel a) of Fig.~\ref{fig2}, where the two pairs or
    points from panels 1 and 2 are highlighted together with their
    corresponding distances.}
  \label{fig1}
\end{figure}
Using
\begin{equation}
\label{eq68}
u_x(x,y)=\sqrt{\frac{2}{\pi}}\,x\,e^{-(x^2+y^2)/2},
\qquad
u_y(x,y)=\sqrt{\frac{2}{\pi}}\,y\,e^{-(x^2+y^2)/2},
\end{equation}
this becomes
\begin{equation}
\label{eq69}
\widetilde\rho^{(2)}(\mathbf r_1,\mathbf r_2) = \frac{4e^{-(r_1^2+r_2^2)}\left[C_{20}\,x_1^2x_2^2 + C_{02}\,y_1^2y_2^2 + C_{11}(x_1y_2+y_1x_2)^2\right]}{\pi^2\left(C_{20}+2C_{11}+C_{02}\right)},
\end{equation}
where \(r_j^2=x_j^2+y_j^2\). Unlike the vortex case, the two-body density is not rotationally invariant. To evaluate the distance distribution, it is convenient to introduce \( \mathbf R\equiv(\mathbf r_1+\mathbf r_2)/2\), \(\mathbf s\equiv\mathbf r_1-\mathbf r_2\), and \(d\equiv|\mathbf s|\) with \(\mathbf R=(X,Y)^T\) and \(\mathbf s=(s_x,s_y)^T\). Then \(r_1^2+r_2^2=2(X^2+Y^2)+d^2/2\)
while the factors read 
\begin{subequations}
  \label{eq:FriApr17045006PMCEST2026}
  \begin{align}
    x_1^2x_2^2 &= \left(X^2-\frac{s_x^2}{4}\right)^2,\\
    y_1^2y_2^2 &= \left(Y^2-\frac{s_y^2}{4}\right)^2,\\
    (x_1y_2+y_1x_2)^2 &= \left(2XY-\frac{s_xs_y}{2}\right)^2.    
  \end{align}
\end{subequations}
Substituting Eq.~\eqref{eq69} into the definition of \(D(d)\), and
carrying out the Gaussian integrals over \(X\) and \(Y\), followed by
the angular average over the orientation of \(\mathbf s\), yields the
closed form
\begin{equation}
\label{eq70}
D(d) = \frac{d}{32}\left[(3-2\mathscr C)(8+d^4)-8(1-2\mathscr C)d^2\right] e^{-d^2/2}\,.
\end{equation}

\begin{figure}[htbp]
  \centering
  \includegraphics[width=\textwidth]{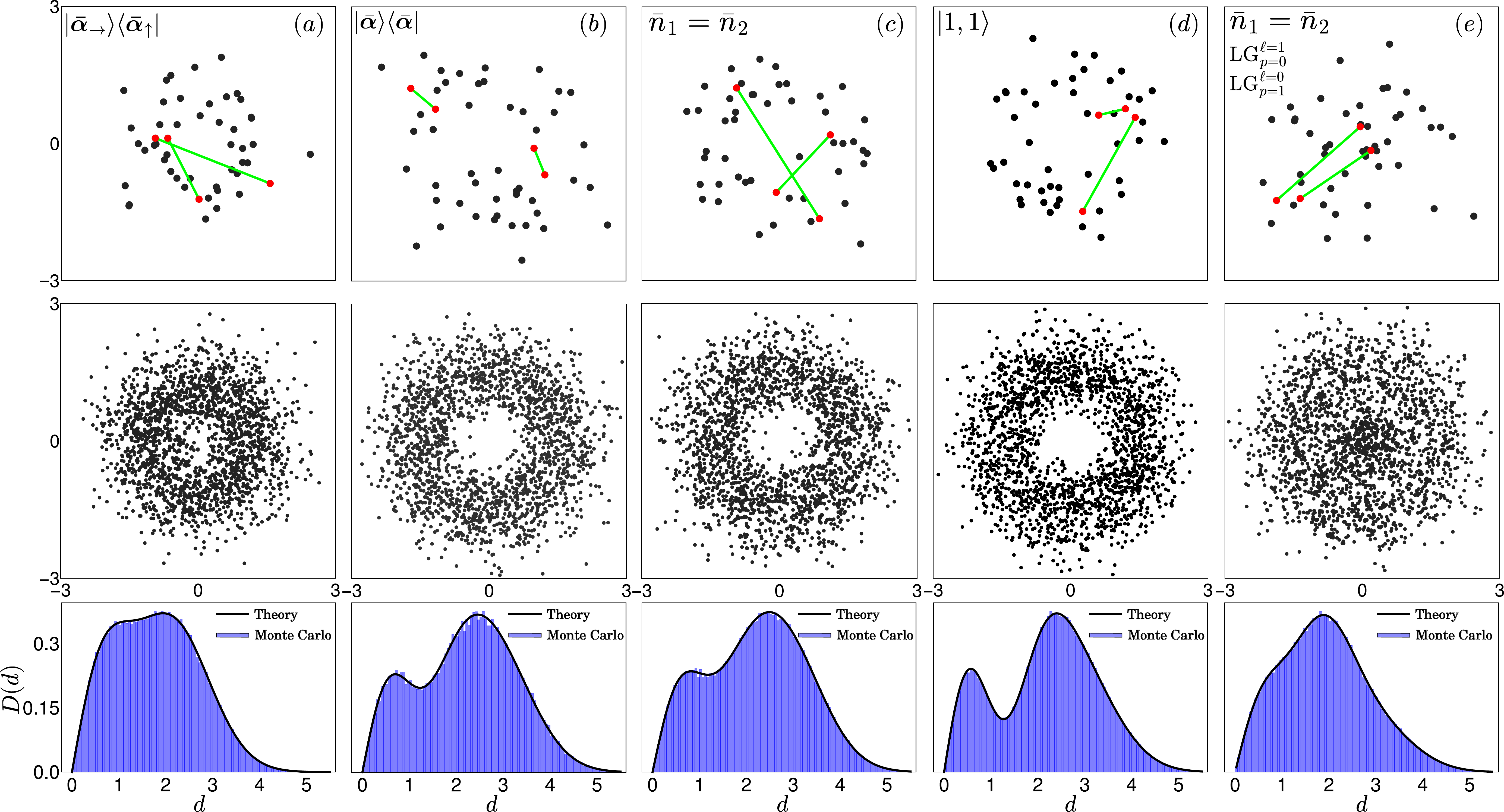}
  \caption{Superframes and corresponding distance distributions for
    the five cases shown in the \(5\times5\) grids of
    Figs.~\ref{fig3}, \ref{fig4}, \ref{fig5}, \ref{fig6}, and
    \ref{fig1}. The upper row shows, in panels a)--e), the
    superposition of all sampled point pairs from the corresponding
    grids, namely: quadrupolar RPCS (Fig.~\ref{fig3}), quadrupolar
    balanced thermal (Fig.~\ref{fig4}(, quadrupolar Fock \(\ket{1,1}\)
    (Fig.~\ref{fig5}), balanced thermal interference of
    LG$_{p=0}^{\ell=1}$ and LG$_{p=1}^{\ell=0}$ (Fig.~\ref{fig6}), and
    dipolar RPCS (Fig.~\ref{fig1}). In each frame, the point pairs
    taken from panels 1 and 2 of the corresponding \(5\times5\) grid
    are highlighted in red, with their mutual distances shown in
    green. The lower row shows the associated distance distributions
    \(D(d)\), obtained from all sampled pairs in each case.}
  \label{fig2}
\end{figure}

Although \(\widetilde\rho^{(2)}(\mathbf r_1,\mathbf r_2)\) is generically anisotropic in the dipolar basis, Eq.~\eqref{eq69} shows that the rotationally invariant observable \(D(d)\) is again fully controlled by the same correlation parameter \(\mathscr C\). In particular, for independent particles sampled from a fixed dipole profile, \(\mathscr C=0\), and one finds
\begin{equation}
\label{eq71}
D_{\mathrm{ind}}^{\rightarrow\uparrow}(d) = \frac{d}{32}\left(24-8d^2+3d^4\right)e^{-d^2/2}.
\end{equation}
On the other hand, for the two-photon Fock state \(\ket{1_x,1_y}\), for which \(\mathscr C=1/2\), Eq.~\eqref{eq70} reduces to
\begin{equation}
\label{eq72}
D_{\ket{1_x,1_y}}(d) = \frac{d}{16}(8+d^4)e^{-d^2/2},
\end{equation}
namely the same functional form as the independent donut
distribution. Our picture remains valid regardless of the choice of
basis.  As in the vortex case, the same observable may be written in
the \(P\) representation as
\begin{equation}
\label{eq73}
D(d) = \frac{\int d^2\beta_x\,d^2\beta_y\, P(\beta_x,\beta_y)\,n_\beta^2\,D_\beta(d)}{\int d^2\beta_x\,d^2\beta_y\, P(\beta_x,\beta_y)\,n_\beta^2},
\end{equation}
where
\begin{equation}
\label{eq74}
D_\beta(d) = \int d^2\mathbf r_1\,d^2\mathbf r_2\, \widetilde\rho_\beta^{(1)}(\mathbf r_1)\,
\widetilde\rho_\beta^{(1)}(\mathbf r_2)\, \delta\!\left(d-|\mathbf r_1-\mathbf r_2|\right).
\end{equation}
This is the form used in the Monte Carlo sampling, where particles are
drawn independently within each shared dipolar geometry
\((\beta_x,\beta_y)\). Fig.~\ref{fig1} illustrates the RPCS state in
the dipole basis through a \(5\times5\) grid of single-shot patterns,
each generated from a geometry
\(\boldsymbol{\beta}\equiv(\beta_x,\beta_y)^T\) drawn from
\(P(\beta_x,\beta_y)\). For each realization, two independently
sampled positions are highlighted in red and their mutual distance is
shown in green. The geometry is less chaotic than for thermal
fluctuations but still varies considerably from shot to shot, as
opposed to the vortex basis where the same structure---a dipole---is
simply rotated.  The corresponding results are summarized in column a)
of Fig.~\ref{fig2}: the top panel shows the frame obtained by
superposing all sampled point pairs from the \(5\times5\) grid, with
the pairs from panels 1 and 2 of Fig.~\ref{fig1} highlighted together
with their distances; the middle panel shows the average spatial
pattern obtained by accumulating many single-shot realizations; and
the bottom panel displays the resulting distance distribution
\(D(d)\).

\section{Still other configurations}

\subsection{Higher~$\ell$ cases}

\begin{figure}[htbp]
  \centering
  \includegraphics[width=0.6\textwidth]{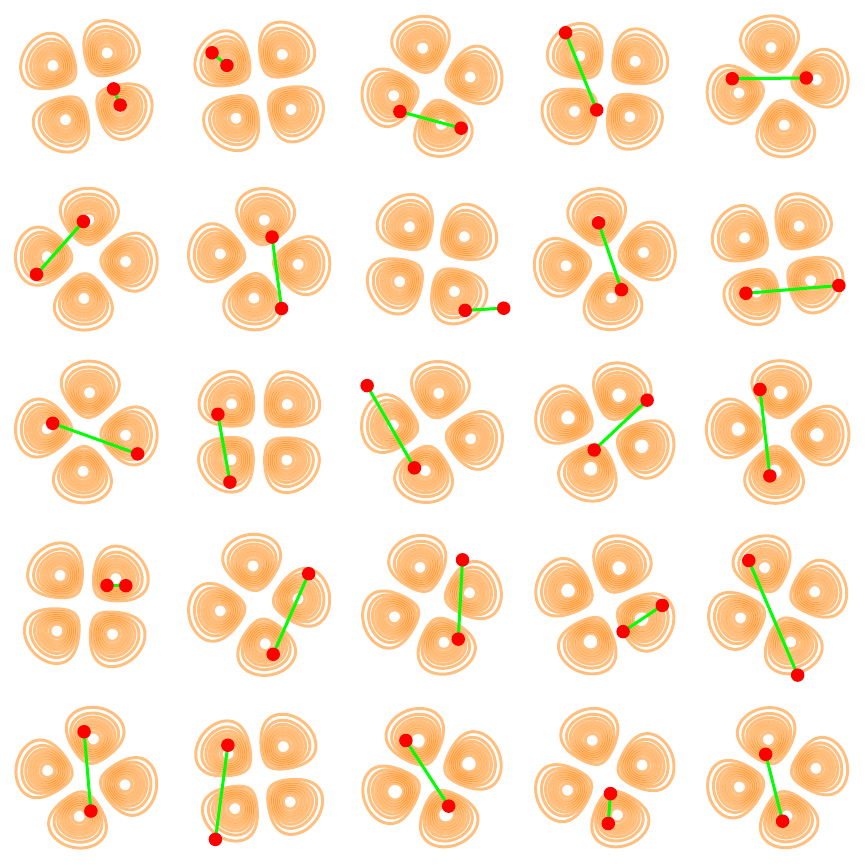}
  \caption{A \(5\times5\) grid of single-shot patterns for the RPCS state in the quadrupolar vortex basis \(\ell=\pm2\). Each panel shows a geometry drawn from the corresponding distribution, together with two independently sampled positions marked in red and their mutual distance shown in green. The superposition of the sampled point pairs from all panels, viewed as a single aggregate frame, is shown in panel b) of Fig.~\ref{fig2}.}
  \label{fig3}
\end{figure}

\begin{figure}[htbp]
  \centering
  \includegraphics[width=0.6\textwidth]{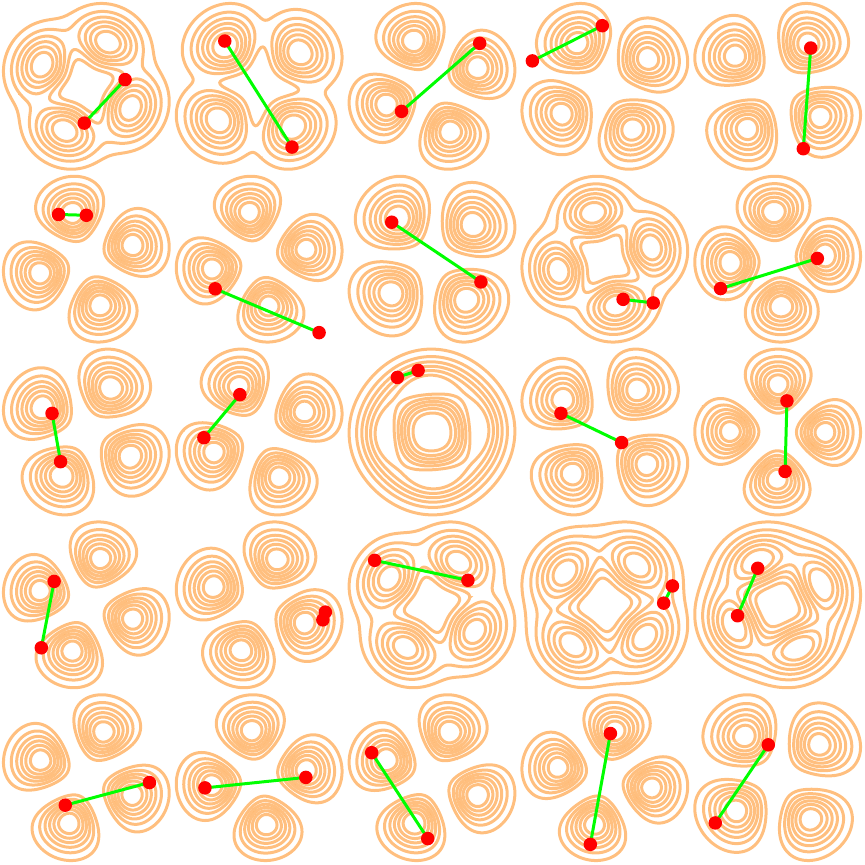}
  \caption{A \(5\times5\) grid of single-shot patterns for the balanced thermal state in the quadrupolar vortex basis \(\ell=\pm2\). Each panel shows a realization of the underlying geometry, together with two independently sampled positions marked in red and their mutual distance shown in green. The superposition of the sampled point pairs from all panels, viewed as a single aggregate frame, is shown in panel c) of Fig.~\ref{fig2}.}
  \label{fig4}
\end{figure}

\begin{figure}[htbp]
  \centering
  \includegraphics[width=0.6\textwidth]{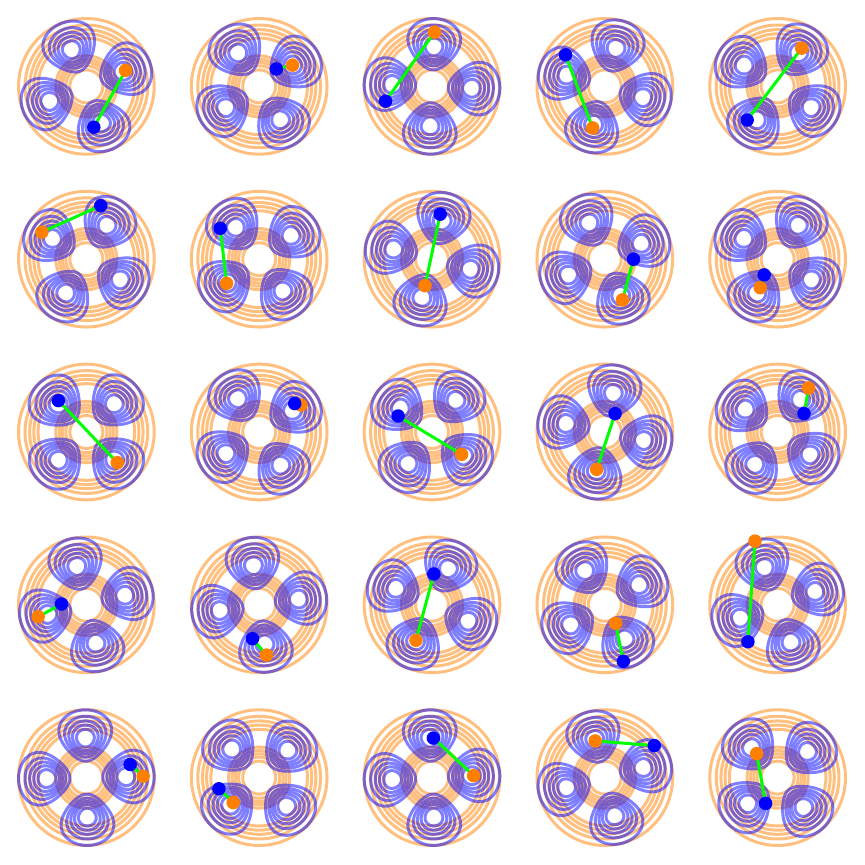}
  \caption{A \(5\times5\) grid of single-shot patterns for the Fock state \(\ket{1,1}\) in the quadrupolar vortex basis \(\ell=\pm2\). In each panel, the first position is sampled from the one-particle density matrix and the second from the corresponding conditional two-particle density matrix; both sampled positions are marked in red and their mutual distance is shown in green. The superposition of the sampled point pairs from all panels, viewed as a single aggregate frame, is shown in panel d) of Fig.~\ref{fig2}.}
  \label{fig5}
\end{figure}

We have focused on~$\ell=\pm1$ in the main text and added above the
lower-excitation cases in the dipole basis. Here, we briefly explore
higher values of~$\ell$, which lead to interferences forming
multipoles in each symmetry breaking. For instance, the case of
quadrupoles~$\ell=\pm2$ is shown in Fig.~\ref{fig3} for for RPCS, in
Fig.~\ref{fig4} for thermal states and in Fig.~\ref{fig5} for Fock
states. As in Fig.~\ref{fig1}, each case is displayed as a
\(5\times5\) grid of single-shot realizations. In every panel, two
sampled positions are marked in red and their distance is shown in
green; in the Fock case, the second position is sampled conditionally
on the first through the corresponding two-particle density. The
corresponding results are shown in columns b), c), and d) of
Fig.~\ref{fig2}: the top row displays the frames obtained by
superposing all sampled point pairs from the respective grids, with
the pairs from panels 1 and 2 highlighted together with their
distances; the middle row shows the average spatial patterns obtained
by accumulating many single-shot realizations; and the bottom row
shows the associated distance distributions \(D(d)\).

There is an obvious connection to the \(\ell=\pm1\) vortex
interferences, the only modifications fro LG$_{p=0}^{\ell=\pm2}$ being
that the angular interference term becomes \(\cos(4\Delta\theta)\) and
the radial profile changes according to
\(|R_2(r)|^2=(r^4 e^{-r^2})/(2\pi)\). Accordingly, the normalized
two-particle reduced density matrix reads
\begin{equation}
\label{eq75}
\widetilde\rho^{(2)}(\mathbf r_1,\mathbf r_2) = \frac{r_1^4 r_2^4 e^{-r_1^2-r_2^2}}{4\pi^2}
\Bigl[1+2\mathscr C\cos(4\Delta\theta)\Bigr],
\end{equation}
where, as before, \(\mathscr C\equiv
C_{11}/(C_{20}+2C_{11}+C_{02})\). The corresponding distance
distribution is then
\begin{equation}
\label{eq76}
D_{\ell=2}(d)
=
\frac{1}{\pi}
\int_0^\infty dr_1\, r_1^5 e^{-r_1^2}
\int_0^\infty dr_2\, r_2^5 e^{-r_2^2}
\int_0^\pi d\Delta\theta\,
\Bigl[1+2\mathscr C\cos(4\Delta\theta)\Bigr]
\delta\!\left(d-\sqrt{r_1^2+r_2^2-2r_1r_2\cos\Delta\theta}\right).
\end{equation}
Carrying out the integrations yields
\begin{equation}
\label{eq77}
D_{\ell=2}(d)
=
\frac{d}{1024}
\Bigl[
d^8+32d^4+384
+
2\mathscr C\bigl(d^8-32d^6+288d^4-768d^2+384\bigr)
\Bigr]
e^{-d^2/2}.
\end{equation}
Equation~\eqref{eq77} is the direct analogue of the \(\ell=\pm1\)
result: the structure of the distance distribution remains fully
controlled by the same correlation parameter \(\mathscr C\).

\subsection{Non-polar symmetric case}
\begin{figure}[htbp]
  \centering
  \includegraphics[width=0.8\textwidth]{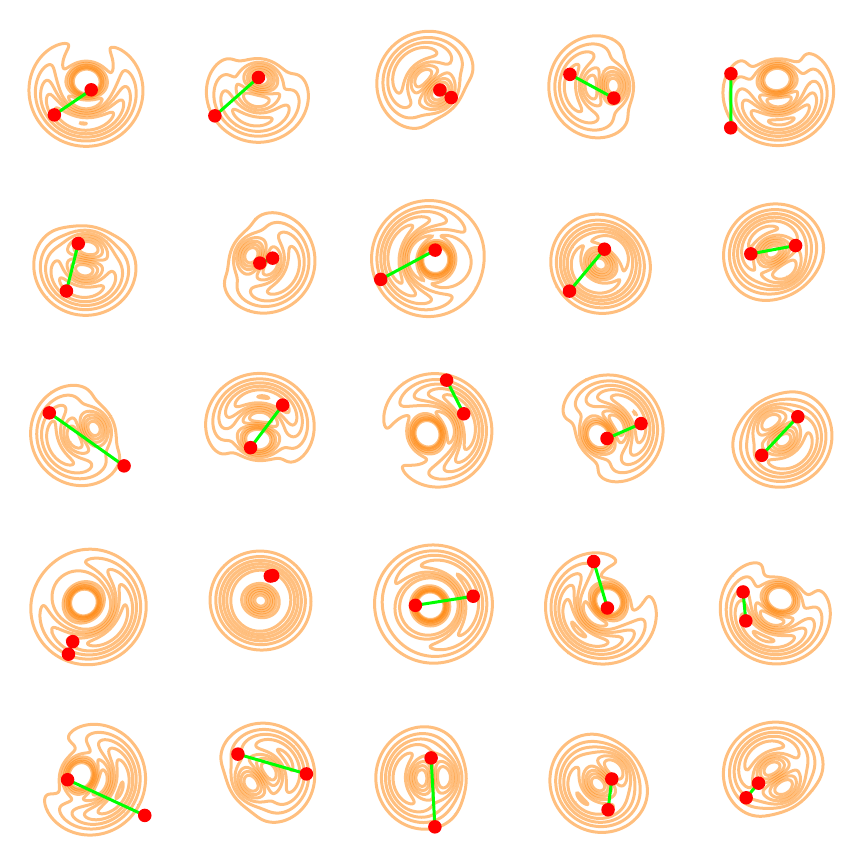}
  \caption{A \(5\times5\) grid of single-shot patterns obtained by interfering the LG$_{p=0}^{\ell=1}$ and LG$_{p=1}^{\ell=0}$ modes, both prepared in balanced thermal states. In each panel, two independently sampled positions are marked in red and their mutual distance is shown in green. 
    The superposition of the sampled point pairs from all panels,
    viewed as a single aggregate frame, is shown in panel e) of
    Fig.~\ref{fig2}.}
  \label{fig6}
\end{figure}

In the~LG$^{\ell=\pm k}$ basis, the correlations are only in the
angular part, and the sampling can thus be made independently for the
radii, regardless of genuine or construed correlations for the
angles. That is to say, Fock states are genuinely correlated in their
angle but their radial part remains independent. This is only a
feature of this particular, polar symmetric geometry, though. In
Fig.~\ref{fig6}, we show a case where both the phases and distances
from the center interfere, namely, LG$_{p=0}^{\ell=1}$ for one mode
and LG$_{p=1}^{\ell=0}$ for the other mode. As a result of this
interplay of complex geometries, complex correlations emerge from the
interplay of phases and distances that remain, however, explainable in
terms of independent, uncorrelated samplings.

As an example, we consider the orthonormal modes
\begin{equation}
\label{eq78}
\phi_a(r,\theta)=\phi_{p=0}^{\ell=1}(r,\theta),
\qquad
\phi_b(r,\theta)=\phi_{p=1}^{\ell=0}(r,\theta),
\end{equation}
which in dimensionless units read
\begin{equation}
\label{eq79}
\phi_a(r,\theta)=\frac{r}{\sqrt{\pi}}e^{-r^2/2}e^{i\theta},
\qquad
\phi_b(r,\theta)=\frac{1-r^2}{\sqrt{\pi}}e^{-r^2/2}.
\end{equation}
Unlike the counter-rotating vortex, the interference term now couples radial and angular dependence through the factor \(r(1-r^2)e^{\pm i\theta}\), so the conditioned one-particle density is no longer separable into purely radial and angular parts. Nevertheless, the \(P\)-representation retains exactly the same structure as before. From Eq.~\eqref{eq17},
\begin{equation}
\label{eq80}
\rho^{(N)}(\mathbf r_1,\dots,\mathbf r_N) = \int d^2\alpha_1\,d^2\alpha_2\, P(\alpha_1,\alpha_2) \prod_{j=1}^N \rho^{(1)}_{\alpha_1,\alpha_2}(\mathbf r_j),
\end{equation}
where
\begin{equation}
\label{eq81}
\rho^{(1)}_{\alpha_1,\alpha_2}(r,\theta) = \frac{e^{-r^2}}{\pi} \Big[|\alpha_1|^2r^2 + |\alpha_2|^2(1-r^2)^2 + 2r(1-r^2)\,\mathrm{Re}\!\big(\alpha_1\alpha_2^*e^{i\theta}\big) \Big].
\end{equation}
Since the modes remain orthonormal \(n_\alpha=|\alpha_1|^2+|\alpha_2|^2\), and the normalized conditioned one-particle density is
\begin{equation}
\label{eq82}
\widetilde\rho^{(1)}_{\alpha_1,\alpha_2}(r,\theta) = \frac{e^{-r^2}}{\pi}\frac{|\alpha_1|^2r^2 + |\alpha_2|^2(1-r^2)^2 + 2r(1-r^2)\,\mathrm{Re}\!\big(\alpha_1\alpha_2^*e^{i\theta}\big)}{|\alpha_1|^2+|\alpha_2|^2}.
\end{equation}
Hence,
\begin{equation}
\label{eq83}
\widetilde\rho^{(N)}(\mathbf r_1,\dots,\mathbf r_N) = \frac{\int d^2\alpha_1\,d^2\alpha_2\, P(\alpha_1,\alpha_2)\, n_\alpha^N \prod_{j=1}^N \widetilde\rho^{(1)}_{\alpha_1,\alpha_2}(\mathbf r_j)}{\int d^2\alpha_1\,d^2\alpha_2\, P(\alpha_1,\alpha_2)\, n_\alpha^N}.
\end{equation}

Thus, the absence of polar symmetry modifies only the explicit form of
the conditioned one-particle density matrix, not the statistical
structure of the many-body state, the latter remains an aggregate over
differing geometries \((\alpha_1,\alpha_2)\), within which particle
coordinates are independent. The balanced and imbalanced thermal cases
therefore follow the same normalization logic as in the vortex basis.

\subsection{Pair-distance distribution for a non-polar-symmetric two-mode basis}\label{sec:distance_nonpolar}

For the same two mode introduced in the previous section, and for states diagonal in the associated Fock basis, the unnormalized two-particle reduced density matrix reads
\begin{equation}
\label{eq84}
\rho^{(2)}(\mathbf r_1,\mathbf r_2) = C_{20}\,|\phi_a(\mathbf r_1)|^2|\phi_a(\mathbf r_2)|^2 + C_{02}\,|\phi_b(\mathbf r_1)|^2|\phi_b(\mathbf r_2)|^2 + C_{11}\,\bigl|\phi_a(\mathbf r_1)\phi_b(\mathbf r_2)+\phi_b(\mathbf r_1)\phi_a(\mathbf r_2)\bigr|^2.
\end{equation}

Using the explicit mode functions, this becomes
\begin{multline}
\label{eq85}
\rho^{(2)}(\mathbf r_1,\mathbf r_2) = \frac{e^{-(r_1^2+r_2^2)}}{\pi^2} \Big[C_{20}\,r_1^2r_2^2 + C_{02}\,(1-r_1^2)^2(1-r_2^2)^2
\\
+ C_{11}\Big(r_1^2(1-r_2^2)^2 + r_2^2(1-r_1^2)^2 + 2r_1r_2(1-r_1^2)(1-r_2^2)\cos\Delta\theta\Big)\Big],
\end{multline}
with \(\Delta\theta=\theta_2-\theta_1\). The normalization is \(\mathcal Z_2=C_{20}+2C_{11}+C_{02}\), so that \(\widetilde\rho^{(2)}(\mathbf r_1,\mathbf r_2)=\rho^{(2)}(\mathbf r_1,\mathbf r_2)/\mathcal {Z}_2\). The corresponding distance distribution is given by Eq.~(\ref{eq57}). Since Eq.~\eqref{eq85} depends only on the relative angle, one obtains
\begin{multline}
\label{eq86}
D(d) = \frac{2}{\pi \mathcal Z_2} \int_0^\infty dr_1\,r_1 e^{-r_1^2} \int_0^\infty dr_2\,r_2 e^{-r_2^2} \int_0^{2\pi} d\Delta\theta\,
\\
\times \Big[C_{20}\,r_1^2r_2^2 + C_{02}\,(1-r_1^2)^2(1-r_2^2)^2 + C_{11}\Big(r_1^2(1-r_2^2)^2 +
r_2^2(1-r_1^2)^2 + 2r_1r_2(1-r_1^2)(1-r_2^2)\cos\Delta\theta\Big)\Big]
\\ \times \delta\!\left(d-\sqrt{r_1^2+r_2^2-2r_1r_2\cos\Delta\theta}\right).
\end{multline}
This is the exact analogue of the vortex expression, with the only difference that the radial kernel is now determined by the non-polar-symmetric mode pair. This behavior is summarized in Fig.~\ref{fig2}(e): the top panel shows the frame obtained by superposing the sampled point pairs from the \(5\times5\) grid of Fig.~\ref{fig6}, the middle panel shows the corresponding average spatial pattern obtained from many realizations, and the bottom panel displays the resulting distance distribution \(D(d)\). In contrast to the previous cases, the averaged pattern in Fig.~\ref{fig2}(e) is no longer donut-like. Instead, it approaches a filled circular profile. This follows directly from the fact that the two interfering modes are not related by polar symmetry and, in the balanced thermal case, the average over the random relative phase removes the interference term. Indeed, for a balanced thermal state, the averaged one-particle density is
\begin{equation}
\label{eq87}
\overline{\rho}^{(1)}(r,\theta) = \frac{1}{2}|\phi_a(r,\theta)|^2+\frac{1}{2}|\phi_b(r,\theta)|^2 = \frac{e^{-r^2}}{2\pi}\left[r^2+(1-r^2)^2\right]= \frac{e^{-r^2}}{2\pi}\left(1-r^2+r^4\right).
\end{equation}
Since
\begin{equation}
\label{eq88}
1-r^2+r^4 = \left(r^2-\frac{1}{2}\right)^2+\frac{3}{4} > 0 \qquad \forall r,
\end{equation}
the averaged profile has no radial zero and therefore no annular hole. Thus, unlike the vortex and dipolar cases discussed previously, the balanced thermal average of this non-polar-symmetric superposition does not produce a donut, but rather a rotationally symmetric filled distribution, which is precisely what is observed in the accumulated pattern.


\bibliographystylesupp{naturemag}
\bibliographysupp{arXiv2}

\end{document}